\documentclass{aa}  

\usepackage{graphicx}
\usepackage{txfonts}
\usepackage{hyperref}
\usepackage{placeins}
\usepackage{stfloats}

\usepackage[dvipsnames]{xcolor}

\newcommand{\dust}{_{\mathrm{d}}}
\newcommand{\gas}{_{\mathrm{g}}}
\newcommand{\sml}{_{\mathrm{d}}^{\mathrm{small}}}
\newcommand{\bg}{_{\mathrm{d}}^{\mathrm{large}}}

\usepackage{soul} 
\usepackage{cancel} 
\setstcolor{cyan}

\begin{document}


\title{Resurgence of CO in a warm bubble around
accreting protoplanets and its observability}

\titlerunning{CO bubble around accreting protoplanets}


   \author{O. Chrenko\inst{\ref{CUNI}},
          S. Casassus\inst{\ref{UCH},\ref{DO}}
          \and
          R. O. Chametla\inst{\ref{CUNI}}
          }
          
   \institute{Charles University, Faculty of Mathematics and Physics, Astronomical Institute, V Hole\v{s}ovi\v{c}k\'{a}ch 747/2, 180 00 Prague 8, Czech Republic \label{CUNI} \\
   \email{chrenko@sirrah.troja.mff.cuni.cz}
   \and
   Departamento de Astronom\'{ı}a, Universidad de Chile, Casilla 36-D, Santiago, Chile \label{UCH}
   \and
   Data Observatory Foundation, Eliodoro Yáñez 2990, Providencia, Santiago, Chile \label{DO}
   }

   \date{Received 18 March 2025 / Accepted 23 June 2025}

\abstract 
{The cold outer regions of protoplanetary disks are expected to contain a midplane-centered layer where gas-phase CO molecules freeze out and their overall abundance is low. The layer then manifests itself as a void in the channel
maps of CO rotational emission lines.
We explore whether the frozen-out layer can expose the circumplanetary
environment of embedded accreting protoplanets to observations.
To this end, we performed 3D radiative gas-dust hydrodynamic simulations with opacities determined by the redistribution of submicron- and millimeter-sized dust grains.
A Jupiter-mass planet with an accretion luminosity of $\sim$$10^{-3}\,L_{\odot}$ was considered as the nominal case.
The accretion heating sustains a warm bubble around the planet, which locally increases the abundance of gas-phase CO molecules.
Radiative transfer predictions of the emergent sky images show that the bubble becomes a conspicuous CO emission source in channel maps. It appears as a low-intensity optically thick spot located in between the
so-called dragonfly wings that trace the fore- and backside line-forming surfaces. The emission intensity of the bubble is nearly independent of the tracing isotopolog, suggesting a very rich observable chemistry, as long as its signal can be deblended from the
extended disk emission. This can be achieved with isotopologs that are optically thin or weakly thermally stratified across the planet-induced gap, such as C$^{18}$O. For these, the bubble stands out as the brightest residual in synthetic ALMA observations after 
subtraction of axially averaged channel maps 
inferred from the disk kinematics, enabling new automatic detections of forming protoplanets.
By contrast, the horseshoe flow steadily depletes
large dust grains from the circumplanetary environment, which becomes
unobservable in the submillimeter continuum,
in accordance with the scarcity of ALMA detections.}


   \maketitle
%

\section{Introduction}
\label{sec:intro}

A wealth of substructures discovered in the outer regions\footnote{In practice, the typical innermost working radii at which 
ALMA can resolve substructures are $\sim$$30\,\mathrm{au}$.}
of protoplanetary disks
\citep[see][and references therein]{Andrews_2020ARA&A..58..483A,Bae_etal_2023ASPC..534..423B,Benisty_etal_2023ASPC..534..605B}
poses a challenge for the theory of planet formation. The question is whether some of these substructures
can indirectly trace embedded protoplanets that perturb their natal disk.
For instance, protoplanets that become massive enough to modify the rotation curve
of the surrounding gas can induce pressure barriers for drifting dust grains that
result in dusty rings \citep[e.g.][]{Paardekooper_Mellema_2006A&A...459L..17P,Zhu_etal_2012ApJ...755....6Z,  Dipierro_etal_2016MNRAS.459L...1D,Zhang_etal_2018ApJ...869L..47Z,Lodato_etal_2019MNRAS.486..453L}.
However, pressure barriers can also form by multiple processes
that are not related to the presence of protoplanets \citep[for reviews see][]{Pinilla_Youdin_2017ASSL..445...91P,Bae_etal_2023ASPC..534..423B}.
Observations of substructures are therefore typically not sufficient to decide
whether protoplanets are present. Supporting evidence is needed.

To date, direct detections of protoplanets have been rare. Several candidates were identified
through high-contrast near-infrared imaging \citep{Keppler_etal_2018A&A...617A..44K,Currie_etal_2022NatAs...6..751C,Hammond_etal_2023MNRAS.522L..51H,Wagner_etal_2023NatAs...7.1208W,Christiaens_etal_2024A&A...685L...1C}. Of these sources, only PDS 70 b and c and AB Aur b exhibit
detectable H$\alpha$ emission \citep{Wagner_etal_2018ApJ...863L...8W,Haffert_etal_2019NatAs...3..749H,Zhou_etal_2021AJ....161..244Z,Zhou_etal_2025ApJ...980L..39Z},
which provides evidence for ongoing gas accretion onto hypothetic giant protoplanets.
Finally, only PDS 70 c can be detected in the submillimeter (submm) continuum \citep{Isella_etal_2019ApJ...879L..25I,Benisty_etal_2021ApJ...916L...2B}, which
indicates a dusty circumplanetary disk (CPD) or free-free emission that is driven by gas accretion \citep[given the variability of the signal;][]{Casassus_Carcamo_2022MNRAS.513.5790C}.
The scarcity of ALMA (Atacama Large Millimeter/submillimeter Array) continuum detections of CPDs  contrasts with existing predictions from isothermal and nonisothermal simulations
\citep[e.g.][]{Wolf_DAngelo_2005ApJ...619.1114W,Szulagyi_etal_2018MNRAS.473.3573S} or steady-state analytical models \citep[][]{zhu2018MNRAS.479.1850Z},  which suggest the formation of a conspicuous and bright CPD in the continuum.

The disk kinematics inferred from submm molecular line emission provide a promising way of detecting protoplanets. Planet-induced perturbations modify
the trajectories of gas parcels and their respective line-of-sight velocities
compared to a disk in pure sub-Keplerian rotation. If the deviations become
strong enough, wiggles or kinks can appear in the channel
maps \citep{Perez_etal_2015ApJ...811L...5P,Pinte_etal_2018ApJ...860L..13P,Pinte_etal_2019NatAs...3.1109P}, which are most commonly obtained
for CO isotopologs.
This kinematic analysis can also be performed based 
on the velocity centroid maps. The idea is to subtract
the axially symmetric background flow and isolate
the velocity reversal in the planet-induced spiral wakes \citep{Perez_etal_2018ApJ...869L..50P,Casassus_Perez_2019ApJ...883L..41C,Chen_Dong_2024ApJ...976...49C}.

Large-scale kinematic features are generated at line-forming surfaces at altitudes of up to a few scale heights \citep[e.g.][]{Law_etal_2022ApJ...932..114L}, offset from the midplane-positioned protoplanets themselves. While these features are relatively ubiquitous in $^{12}$CO, however, they are surprisingly less pronounced in rarer isotopologs that probe deeper disk layers \citep{Perez_etal_2020ApJ...889L..24P}, and their origin cannot be uniquely attributed to a specific body \citep[e.g. the case of AS\,209;][]{Bae_etal_2022ApJ...934L..20B,Fedele_etal_2023A&A...672A.125F}. Interestingly, the models that yield submm-bright CPDs also predict bright molecular line emission that is expected to persist in multiple channels \citep[][]{Perez_etal_2015ApJ...811L...5P}. Nevertheless, the line persistence effect has not yet been convincingly detected \citep[except possibly in AS\,209 and Elias\,2-24;][]{Bae_etal_2022ApJ...934L..20B,Pinte_etal_2023MNRAS.526L..41P}.

Here, we examine the tension between the observed wealth of structure in disks and the scarcity of radio-continuum protoplanet detections, and we consider whether the accretion luminosity of an embedded protoplanet can facilitate a detectable and rich chemistry of the circumplanetary environment regardless of the formation of a CPD \citep[see also][]{Cleeves_etal_2015ApJ...807....2C,Jiang_etal_2023A&A...678A..33J,Petrovic_etal_2024MNRAS.534.2412P}.
We use the fact that outside the CO snowline\footnote{The location of the CO snowline can vary from one disk to the next; as an example. we quote $\sim$$30\,\mathrm{au}$ for TW Hya \citep{Qi_etal_2013Sci...341..630Q} and $\sim$$80\,\mathrm{au}$ for HD 163296 \citep{Qi_Wilner_2024ApJ...977...60Q}.}, the gas-phase molecules
of CO freeze out and become depleted in an interior layer of the disk.
Consequently, for moderately inclined disks, the fore- and backside
`dragonfly wings' that appear in channel maps become separated by a void from which CO is absent
\citep[for similar considerations, see][]{Dullemond_etal_2020A&A...633A.137D}.
We envisage that if the accreting luminous protoplanet 
orbits within the CO-depleted layer,
it heats up the surrounding gas \citep{Benitez-Llambay_etal_2015Natur.520...63B,Montesinos_etal_2015ApJ...806..253M,Szulagyi_2017ApJ...842..103S,Muley_etal_2024A&A...687A.213M} and
prevents freeze-out locally, which means that the abundance of CO molecules is higher than in the cold surroundings.
When the resulting CO bubble that encapsulates the planet is observed at a convenient viewing geometry,
it might be localized within the emission void and might also be isolated as a kinematic signal.

To provide a proof of concept, we performed 3D dust-gas radiation hydrodynamic simulations of disk-embedded luminous protoplanets, converted the result into synthetic images, and searched for observable features.  Important aspects of our modeling involved the direct treatment of heating and cooling processes, which is essential to avoid ambiguity between the temperature profile and the dynamical state of the gas, and direct tracking of millimeter-sized (mm-sized) dust grains, which allowed us to predict the thermal continuum emission together with the molecular line emission. Our hydrodynamic and radiative transfer methods are described in Sect.\,\ref{sec:methods}. A nominal simulation, described in Sect.\,\ref{sec:nominal}, serves as a case study in conditions whose observability is discussed in Sect.\,\ref{sec:obs}. 
The supporting analysis of the nominal simulation is provided in Appendix~\ref{sec:dust_depletion}, and
additional configurations are documented in Appendix~\ref{sec:additional}.
We discuss our findings and conclude in Sect.\,\ref{sec:conclusions}.

\section{Methods} \label{sec:methods}

\subsection{Equations of 3D two-fluid radiation hydrodynamics}
\label{sec:equations}

We modeled a 3D protoplanetary disk
on an Eulerian mesh in spherical coordinates (radius $r$, azimuth $\phi$, and colatitude $\theta$).
The disk comprised three components: gas, large dust grains
(with a physical size $a\bg=1\,\mathrm{mm}$), and small dust grains ($a\sml=0.1\,\mu\mathrm{m}$). The two-population
dust model follows the reasoning of \cite{Ziampras_etal_2025MNRAS.536.3322Z} and allowed us to (i) study the effect of the redistribution of large grains on the thermal properties of the disk, (ii) predict dust continuum observations, and (iii) account for dust coagulation at 
least in a simple parametric manner.
To do this, we introduced a coagulation parameter that sets the initial solid mass fraction contained
in mm-sized grains,
\begin{equation}
    X_{0} = \frac{\Sigma\bg}{\Sigma\bg+\Sigma\sml}\bigg\rvert_{t=0} \equiv 0.9 \, ,
    \label{eq:x0}
\end{equation}
where $\Sigma\bg$ and $\Sigma\sml$ are the surface densities
of large and small dust grains, respectively\footnote{Our choice 
of $X_{0}$ represents a moderate value and is motivated by the fact 
that (i) a substantially lower $X_{0}$ would lead to absorption
of stellar irradiation already at very high elevations above midplane, resulting
in a vertically extended cold disk interior, and hence, in an extremely extended
zone of frozen-out CO; 
(ii) a substantially higher $X_{0}$ would lead to extremely low optical
depths in regions that are devoid of large grains,
such as within planet-induced gaps. We wished to avoid both (i) and (ii).}.

The population of small grains was strongly aerodynamically coupled with the gas. We assumed that the coupling was ideal so that the small grains behaved as tracers of the gas distribution
and did not have to be evolved explicitly. The volume density of small dust grains $\rho\sml$ can be derived
from the gas density $\rho\gas$ at any time using
\begin{equation}
    \rho\sml = (1-X_{0})Z_{0}\rho\gas \, ,
    \label{eq:rho_small}
\end{equation}
where $Z_{0}=\Sigma\dust/\Sigma\gas|_{t=0}=(\Sigma\bg+\Sigma\sml)/\Sigma\gas|_{t=0}\equiv0.01$ is the initial dust-to-gas ratio.

To model the coupled evolution of gas and large grains, we adopted
a pressureless fluid approximation for the latter and solved the following equations of two-fluid hydrodynamics:
\begin{equation}
    \frac{\partial\rho_{\mathrm{g}}}{\partial t}+\nabla\cdot\left(\rho_{\mathrm{g}}\vec{v}\right) = 0 \, ,
    \label{eq:conti_g}
\end{equation}
\begin{equation}
    \frac{\partial\vec{v}}{\partial t}+\left(\vec{v}\cdot\nabla\right)\vec{v} = -\frac{1}{\rho_{\mathrm{g}}}\nabla P + \frac{1}{\rho_{\mathrm{g}}}\nabla\cdot\tens{T}-\nabla\Phi-\frac{\rho_{\mathrm{d}}^{\mathrm{large}}}{\rho_{\mathrm{g}}}\vec{a}_{\mathrm{drag}} \, ,
    \label{eq:navier_g}
\end{equation}
\begin{equation}
    \frac{\partial\rho_{\mathrm{d}}^{\mathrm{large}}}{\partial t}+\nabla\cdot\left(\rho_{\mathrm{d}}^{\mathrm{large}}\vec{u} + \vec{j}\right) = 0 \, ,
    \label{eq:conti_d}
\end{equation}
\begin{equation}
    \frac{\partial\vec{u}}{\partial t}+\left(\vec{u}\cdot\nabla\right)\vec{u} = -\nabla\Phi + \vec{a}_{\mathrm{drag}} \, ,
    \label{eq:navier_d}
\end{equation}
where we introduced the gas velocity $\vec{v}$, the pressure $P$,
the viscous stress tensor $\tens{T}$, the gravitational potential $\Phi$,
the aerodynamic drag acceleration $\vec{a}_{\mathrm{drag}}$, the dust
velocity $\vec{u}$, and the turbulent dust diffusion flux $\vec{j}$.

The equation of state was that of an ideal gas,
\begin{equation}
    P = (\gamma-1)\epsilon = (\gamma-1)\rho\gas c_{V} T \, ,
    \label{eq:eos}
\end{equation}
where $\gamma=1.43$ is the adiabatic index, $\epsilon$ is the internal energy density of gas,
$c_{V}$ is the heat capacity at constant volume, and $T$ is the temperature.
We assumed that the gas and dust grains were thermalized and therefore
described by the same $T$.

The kinematic gas viscosity was parameterized following \cite{Shakura_Sunyaev_1973A&A....24..337S},
\begin{equation}
\nu = \alpha \frac{c_{\mathrm{s}}^{2}}{\Omega_{\mathrm{K}}} \, ,
\label{eq:visco}
\end{equation}
with $\alpha=5\times10^{-4}$, $c_{\mathrm{s}}$ being the sound speed, and $\Omega_{\mathrm{K}}$
being the Keplerian orbital frequency. 
The viscosity enters the calculation through tensor $\tens{T}$
\citep[for a full definition of $\tens{T}$, see][]{Benitez-Llambay_Masset_2016ApJS..223...11B},
and because it is thought to be driven by unresolved underlying turbulence,
we assumed that it sets the level of dust diffusion as well \citep{Morfill_Voelk_1984ApJ...287..371M},
\begin{equation}
    \vec{j} = - \nu(\rho\gas+\rho\bg)\nabla\left(\frac{\rho\bg}{\rho\gas+\rho\bg}\right) \, .
    \label{eq:diffusion}
\end{equation}

The gravitational potential reads
\begin{equation}
    \Phi = -\frac{GM_{\star}}{r} - \frac{GM_{\mathrm{p}}}{d}\left[\left(\frac{d}{r_{\mathrm{sm}}}\right)^{4}-2\left(\frac{d}{r_{\mathrm{sm}}}\right)^{3}+2\frac{d}{r_{\mathrm{sm}}}\right] \,,
    \label{eq:pot}
\end{equation}
where $G$ is the gravitational constant, $M_{\star}$ is the mass of
the central star, $M_{\mathrm{p}}$ is the planet mass, $d$ is the cell-planet distance, and $r_{\mathrm{sm}}$ is the smoothing length. The potential of an embedded planet (the last term in Eq.~\ref{eq:pot}) was 
smoothed with a cubic spline as in \cite{Klahr_Kley_2006A&A...445..747K}.

The aerodynamic drag was considered as a two-way acceleration (it appears
with opposite signs in Eqs.~\ref{eq:navier_g} and \ref{eq:navier_d})
to preserve momentum conservation and was defined as
\begin{equation}
    \vec{a}_{\mathrm{drag}}=\frac{\Omega_{\mathrm{K}}}{\mathrm{St}}\left(\vec{v}-\vec{u}\right) \, ,
    \label{eq:drag}
\end{equation}
where $\mathrm{St}$ is the Stokes number (the dimensionless stopping time) of large dust grains. 
In our model, the Stokes number was set to vary from one cell to the next according to
\begin{equation}
    \mathrm{St} = a\bg\sqrt{\frac{\pi\gamma}{8}}\frac{\rho_{\mathrm{mat}}\Omega_{\mathrm{K}}}{\rho\gas c_{\mathrm{s}}} \, ,
    \label{eq:st}
\end{equation}
where $\rho_{\mathrm{mat}}$ is the material density of large grains. We capped the Stokes number as $\mathrm{St}\leq0.5$, however, in order
to ensure that (i) the Schmidt number\footnote{The Schmidt number is the ratio of the turbulent viscosity $\nu$ to the dust diffusivity \citep{Cuzzi_etal_1993Icar..106..102C}.} remained $\approx$$1$ (which is implicitly assumed in writing Eq.~\ref{eq:diffusion}) and (ii) the fluid approximation for dust remained valid even in disk regions with a very low gas density \citep[see also][]{Krapp_etal_2024ApJ...973..153K}.

The thermodynamics of the disk was regulated via energy equations for the gas
and diffuse thermal radiation
\citep{Commercon_etal_2011A&A...529A..35C,Bitsch_etal_2013A&A...549A.124B}, which read
\begin{equation}
    \frac{\partial\epsilon}{\partial t} + \nabla\cdot\left(\epsilon\vec{v}\right) = - \rho_{\mathrm{g}}\kappa_{\mathrm{P}}\left[4\sigma T^{4} - cE_{\mathrm{R}}\right] - P\nabla\cdot\vec{v} + Q_{+} \, ,
    \label{eq:energy}
\end{equation}
\begin{equation}
    \frac{\partial E_{\mathrm{R}}}{\partial t} + \nabla\cdot\vec{F} = \rho_{\mathrm{g}}\kappa_{\mathrm{P}}\left[4\sigma T^{4} - cE_{\mathrm{R}}\right] \, ,
    \label{eq:energy_rad}
\end{equation}
where $\kappa_{\mathrm{P}}$ is the Planck opacity per gram of gas, $\sigma$
is the Stefan constant, $c$ is the speed of light, $Q_{+}$
is a symbolic notation encapsulating all considered heating terms, $E_{\mathrm{R}}$ is the 
radiative energy density of diffuse thermal radiation, and $\vec{F}$ is the radiation flux vector.
The latter is calculated as
\begin{equation}
    \vec{F} = - \lambda_{\mathrm{lim}}\frac{c}{\rho\gas\kappa_{\mathrm{R}}}\nabla E_{\mathrm{R}} \, ,
    \label{eq:fld}
\end{equation}
where $\lambda_{\mathrm{lim}}$ is the flux limiter \citep{Levermoe_Pomraning_1981ApJ...248..321L,Kley_1989A&A...208...98K}, and
$\kappa_{\mathrm{R}}$ is the Rosseland opacity per gram of gas.

Regarding the heating sources $Q_{+}$, we accounted for viscous heating \citep{Mihalas_WeibelMihalas_1984frh..book.....M}, 
radially ray-traced frequency-averaged stellar irradiation
\citep{Chrenko_Nesvorny_2020A&A...642A.219C}, heating due to the shock-spreading
viscosity used in finite-difference schemes \citep{Stone_Norman_1992ApJS...80..753S},
and heating due to the accretion luminosity of embedded planets \citep{Benitez-Llambay_etal_2015Natur.520...63B,Chrenko_Lambrechts_2019}. The accretion luminosity
was parameterized by the mass doubling time $t_{\mathrm{acc}}$ as
\begin{equation}
    L = \frac{GM_{\mathrm{p}}^{2}}{R_{\mathrm{p}}t_{\mathrm{acc}}} \, ,
    \label{eq:luminosity}
\end{equation}
where $R_{\mathrm{P}}$ is the surface radius of the planet.

Eqs.~(\ref{eq:conti_g})--(\ref{eq:navier_d}) 
were numerically solved using the multifluid version of the code \textsc{Fargo3D}\footnote{https://github.com/FARGO3D/fargo3d}  \citep{Benitez-Llambay_Masset_2016ApJS..223...11B,Benitez-Llambay_2019ApJS..241...25B}
along with an implicit solver for Eqs.~(\ref{eq:energy})--(\ref{eq:energy_rad}) \citep{Chrenko_Lambrechts_2019,Chrenko_Nesvorny_2020A&A...642A.219C}.
The dust diffusion module was taken from \cite{Weber_etal_2019ApJ...884..178W}.

\begin{figure}
    \centering
    \includegraphics[width=0.98\columnwidth]{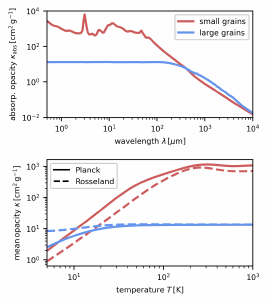}
    \caption{Absorption opacities as a function of the wavelength (\emph{top})
    and mean opacities as a function of the temperature (\emph{bottom})
    for populations of small (red; $a\sml=0.1\,\mu\mathrm{m}$) and large (blue; $a\bg=1\,\mathrm{mm}$) dust grains.
    In the \emph{bottom panel}, the Planck and Rosseland
    opacities are shown
    as solid and dashed curves, respectively.}
    \label{fig:opacity}
\end{figure}

\subsection{Two-population dust opacity}
\label{sec:opacity}

The radiative transfer in the disk is regulated by dust grains\footnote{Gas opacities
were neglected in our model.}. The opacity laws we used in our model are similar
to those of \cite{Ziampras_etal_2025MNRAS.536.3322Z}.
To establish these laws, we first calculated the frequency-dependent opacities of small and large dust grains
using the code \textsc{optool} \citep{Dominik_OPTOOL_2021ascl.soft04010D}.
We adopted the standard DSHARP material composition \citep{Birnstiel_etal_2018ApJ...869L..45B},
but added a moderate porosity of $50\%$ motivated by recent studies \citep{Zhang_2023,Ueda_etal_2024NatAs...8.1148U}. This resulted in $\rho_{\mathrm{mat}}=0.84\,\mathrm{g}\,\mathrm{cm}^{-3}$. The absorption opacities are shown in the top panel of Fig.~\ref{fig:opacity}.

Next, we calculated the frequency-averaged Planck and Rosseland opacities for each grain population. The Planck average was taken over the absorption opacities, and the Rosseland
mean was taken over the extinction opacities. The resulting opacity curves are shown
in the bottom panel of Fig.~\ref{fig:opacity}. They can be approximated with the following
fits:
\begin{equation}
    \kappa_{\mathrm{R}}^{\mathrm{small}} = \min\left[0.031T^{2.05},1.21T^{1.24},746.42\right]\,\mathrm{cm}^{2}\,\mathrm{g}^{-1} \, ,
    \label{eq:kR_small}
\end{equation}
\begin{equation}
    \kappa_{\mathrm{P}}^{\mathrm{small}} = \min\left[0.058T^{2.18},7.82T^{0.92},1069.5\right]\,\mathrm{cm}^{2}\,\mathrm{g}^{-1} \, ,
    \label{eq:kP_small}
\end{equation}
\begin{equation}
    \kappa_{\mathrm{R}}^{\mathrm{large}} = \min\left[5.09T^{0.29},13.74\right]\,\mathrm{cm}^{2}\,\mathrm{g}^{-1} \, ,
    \label{eq:kR_large}
\end{equation}
\begin{equation}
    \kappa_{\mathrm{P}}^{\mathrm{large}} = \min\left[0.76T^{0.85},13.1\right]\,\mathrm{cm}^{2}\,\mathrm{g}^{-1} \, .
    \label{eq:kP_large}
\end{equation}
Additionally, we calculated the opacity
to stellar irradiation as the Planck opacity
at the effective temperature of the irradiating star ($T_{\star}=4500\,\mathrm{K}$ hereinafter), which led to
\begin{equation}
    \kappa_{\star}^{\mathrm{small}} = 1799.38\,\mathrm{cm}^{2}\,\mathrm{g}^{-1} \, ,
    \label{eq:kS_small}
\end{equation}
\begin{equation}
    \kappa_{\star}^{\mathrm{large}} = 13.24\,\mathrm{cm}^{2}\,\mathrm{g}^{-1} \, .
    \label{eq:kS_large}
\end{equation}
The opacities $\kappa_{\star}^{\mathrm{small}}$ and $\kappa_{\star}^{\mathrm{large}}$ determine how the flux of impinging stellar photons is attenuated in the stellar irradiation heating term \citep{Kolb_etal_2013A&A...559A..80K,Chrenko_Nesvorny_2020A&A...642A.219C}.

To obtain $\kappa_{\mathrm{R}}$ and $\kappa_{\mathrm{P}}$ and update them in each grid cell,
we first calculated the local mass fraction of large grains at each time $t$,
\begin{equation}
    X(t) = \frac{\rho\bg}{\rho\bg+\rho\sml} \, ,
\end{equation}
and derived the opacity of the mixture as \citet{Ziampras_etal_2025MNRAS.536.3322Z}
\begin{equation}
    \kappa_{\mathrm{R}/\mathrm{P}/\star} = (1-X)\kappa_{\mathrm{R}/\mathrm{P}/\star}^{\mathrm{small}} + X\kappa_{\mathrm{R}/\mathrm{P}/\star}^{\mathrm{large}} \, .
    \label{eq:kappa_mix}
\end{equation}
We emphasize that the opacities resulting from Eq.~(\ref{eq:kappa_mix}) are
expressed per gram of dust and need to be rescaled per gram of gas before they are used in Eqs.~(\ref{eq:energy})--(\ref{eq:energy_rad}).

\subsection{Simulation stages}
\label{sec:stages}

Our simulations involved three distinct stages.
The first two stages assumed that the disk is
symmetric along its rotational axis (thus becoming effectively 2D)
and allowed us to find an equilibrium between the temperature
structure of the disk and the spatial distribution of the gas and dust.
The final third stage involved the main 3D run with an embedded planet.

\begin{table}[]
    \caption{Summary of the parameters of our nominal hydrodynamic simulation.}
    \centering
    \begin{tabular}{lr}
    \hline\hline
    Grid resolution\tablefootmark{(a)} ($r\times\phi\times\theta$) & $400\times1300\times192$ \\
    Opening angle in colatitude & $50^{\circ}$ \\
    Azimuthal coverage\tablefootmark{(a)} & $360^{\circ}$ \\
    Inner radial boundary\tablefootmark{(a)} & 40 au \\
    Outer radial boundary\tablefootmark{(a)} & 250 au \\
    Planet's orbital distance\tablefootmark{(a)} & $r_{\mathrm{p}}=120\,\mathrm{au}$ \\
    Initial gas surface density & Eq.~(\ref{eq:surfdens}) \\
    Viscosity parameter & $\alpha=5\times10^{-4}$ \\
    Adiabatic index & $\gamma=1.43$ \\
    Mean molecular weight & $\mu=2.3$ \\
    Initial dust-to-gas ratio & $Z_{0}=10^{-2}$ \\
    Initial solid mass fraction in large grains & $X_{0}=0.9$ \\
    Small dust grain size & $a\sml=0.1\,\mu\mathrm{m}$ \\
    Large dust grain size & $a\bg=1\,\mathrm{mm}$ \\
    Dust material density & $\rho_{\mathrm{mat}}=0.84\,\mathrm{g}\,\mathrm{cm}^{-3}$ \\
    Dust opacity & Sect.~\ref{sec:opacity} \\
    Stellar temperature & $T_{\star}=4500\,\mathrm{K}$ \\
    Stellar radius & $R_{\star}=2.5\,R_{\odot}$ \\
    Stellar mass & $M_{\star}=1\,M_{\odot}$ \\
    Planet mass $M_{\mathrm{p}}$\tablefootmark{(a)} & $1\,M_{\mathrm{Jup}}$ \\
    Mass doubling time $t_{\mathrm{acc}}$\tablefootmark{(a)} & $0.1\,\mathrm{Myr}$  \\
    Smoothing length $r_{\mathrm{sm}}$\tablefootmark{(a)} & $0.25\,R_{\mathrm{H}}$  \\
    No. of orbits to introduce $M_{\mathrm{p}}$\tablefootmark{(a)} & 50  \\
    No. of simulated orbits\tablefootmark{(a)} & 800 \\
    \hline
    \end{tabular}
    \tablefoot{\tablefoottext{a}{Applies to the final simulation stage with an embedded planet.}}   
    \label{tab:params_nominal}
\end{table}

\subsubsection{2D hydrostatic relaxation}

In this stage, the domain stretched from 
1 to 300 au in radius and had a total opening
angle of $50^{\circ}$ in colatitude
($25^{\circ}$ per one hemisphere).
We used 1024 logarithmically spaced grid cells in radius and
192 uniformly spaced grid cells in colatitude.
The equilibrium disk structure was searched
iteratively. Starting with a fixed temperature field, we solved the equations
of hydrostatic equilibrium \citep[see e.g.][]{Flock_etal_2016ApJ...827..144F,Chrenko_Nesvorny_2020A&A...642A.219C} that let us derive the distribution of $\rho\gas$.
To solve the hydrostatic equations, a constraint on the radial profile of the surface density has
to be provided. Unless stated otherwise, we considered a smooth power-law disk described by
\begin{equation}
    \Sigma\gas = 100\left(\frac{r}{1\,\mathrm{au}}\right)^{-1/2}\,\mathrm{g}\,\mathrm{cm}^{-2} \, .
    \label{eq:surfdens}
\end{equation}

With the new estimate of $\rho\gas$, we updated the distribution of small and large dust grains. We used Eq.~(\ref{eq:rho_small}) for the former, and the latter was obtained as
\begin{equation}
    \rho\bg = X_{0}Z_{0}\frac{\Sigma\gas}{\sqrt{2\pi}H_{\mathrm{d}}}\exp{\left(-\frac{z^{2}}{2H_{\mathrm{d}}^{2}}\right)} \, ,
    \label{eq:rhodust_hydrost}
\end{equation}
where $z$ is the cylindrical vertical height above the midplane,
and $H_{\mathrm{d}}$ is the scale height of large grains. We followed \cite{Dubrulle_etal_1995Icar..114..237D}
and set
\begin{equation}
    H_{\mathrm{d}} = H\sqrt{\frac{\alpha}{\alpha+\mathrm{St_{2D}}}} \, ,
    \label{eq:Hdust}
\end{equation}
where $H=c_{\mathrm{s}}/(\sqrt{\gamma}\Omega_{\mathrm{K}})$ is the pressure scale height
of gas and 
\begin{equation}
    \mathrm{St_{2D}} = a\bg\frac{\pi}{2}\frac{\rho_{\mathrm{mat}}}{\Sigma\gas} \, ,
    \label{eq:st2d}
\end{equation}
is the vertically integrated form of the Stokes number.

To close one iteration, we kept the gas and dust distributions fixed
and solved Eqs.~(\ref{eq:energy})--(\ref{eq:energy_rad}) over a time interval $\sim$$10^{6}\,\mathrm{s}$
\citep[see also][]{Melon_Fuksman_2022ApJ...936...16M},
thus obtaining an updated temperature profile. We then returned to calculating
$\rho\gas$ until the relative precision between two consecutive iterations dropped below $10^{-5}$.

The boundary conditions we used to solve the equations of hydrostatic equilibrium
were symmetric. In Eqs.~(\ref{eq:energy})--(\ref{eq:energy_rad}), the boundary conditions 
for $\epsilon$ were symmetric as well, whereas $E_{\mathrm{R}}$ was symmetrized only at the inner
radial boundary and set to $E_{\mathrm{R}}=4\sigma/c(5\,\mathrm{K})^{4}$ elsewhere. The latter
condition allows the disk to cool freely by the escape of thermal radiation \citep[see also][]{Chrenko_etal_2024AJ....167..124C}.

\subsubsection{2D hydrodynamic relaxation}

In this stage, the result of the hydrostatic relaxation was remapped to
a grid with a narrower radial span, ranging from 40 to 250 au. The radial
domain was sampled by 400 logarithmically spaced cells.
Subsequently, the full hydrodynamic model introduced in Sect.~\ref{sec:equations}
was evolved over several hundred orbital timescales (the exact simulation time spans
are specified in Sects.~\ref{sec:nominal} and \ref{sec:additional}).

We used the symmetric boundary conditions for $\rho\gas$, $\epsilon$, and $\rho\bg$.
The boundary for $v_{\phi}$ and $u_{\phi}$ was symmetric in colatitude and a Keplerian extrapolation
was used in radius. The boundary for $v_{\theta}$ and $u_{\theta}$ was symmetric in radius, and outflow was allowed in colatitude. The condition for $v_{r}$ was antisymmetric in radius and symmetric in colatitude.
The same was true for $u_{r}$, with the exception of the inner radial boundary, where we allowed for dust outflow. For $E_{\mathrm{R}}$, we again set $E_{\mathrm{R}}=4\sigma/c(5\,\mathrm{K})^{4}$ in colatitude,
but in the radial direction, we fixed $E_{\mathrm{R}}$ to values found during the hydrostatic relaxation
at the respective radial distance. We chose a similar approach to calculate the radial
optical depth $\tau_{\mathrm{irr}}$ to stellar irradiation: Its value at the inner radial edge
was directly informed from the hydrostatic relaxation.

Additionally, the boundaries can be supplemented with buffer zones in which quantities
can be damped following \cite{deValBorro_etal_2006MNRAS.370..529D}. During the hydrodynamic relaxation,
damping was applied near both radial edges, keeping $\rho\gas$ near its hydrostatic equilibrium state and $(v_{r}, v_{\theta})$ near zero.
For the large dust grains, we damped $\rho\bg$ to its hydrostatic equilibrium
state near the outer radial edge and thus created a reservoir that replenished the dust population
by radial drift.

\subsubsection{3D run with an embedded planet}

The main simulation stage was initialized from the result of the hydrodynamic relaxation
by copying it 1300 times in azimuth.
Additionally, the velocity fields were converted into a frame that corotated with the planet.
In our simulations, the orbit of the planet was kept circular and nonmigrating.

The boundary conditions remained the same as during the hydrodynamic relaxation.
In order to improve numerical stability, however, we added an additional
damping zone at the disk surfaces in colatitude. In these zones, we damped all dust-related quantities, $v_{r}$, and $v_{\theta}$ to the values found by hydrodynamic relaxation.
Without this damping, the numerical scheme tends to occasionally
crash, likely because of the extreme dust density contrast at high altitudes above the midplane
(we found minimum dust densities as low as $10^{-180}\,\mathrm{g}\,\mathrm{cm}^{-3}$).
It might be argued that the damping artificially inserts additional dust in the simulation that 
can then settle toward the midplane and affect the solution, but we verified that 
this effect is negligible because the dust densities are very low, as described above.

\begin{figure}
    \centering
    \includegraphics[width=0.98\columnwidth]{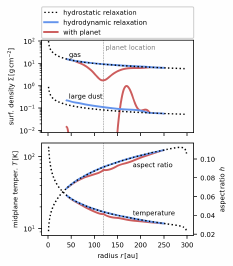}
    \caption{Azimuthally averaged radial profiles in our nominal simulation
    with a Jupiter-mass planet. \emph{Top:} Vertically integrated
    surface density of gas and large dust. \emph{Bottom:} Midplane
    temperature (primary vertical axis) and aspect ratio (secondary vertical axis).
    We show the final state after the hydrostatic relaxation (dotted black curve),
    the hydrodynamic relaxation (solid blue curve), and the main stage with the 
    embedded planet (solid red curve). The planet location is marked with
    the dashed vertical line.}
    \label{fig:M1_profiles}
\end{figure}

\begin{figure}
    \centering
    \includegraphics[width=0.98\linewidth]{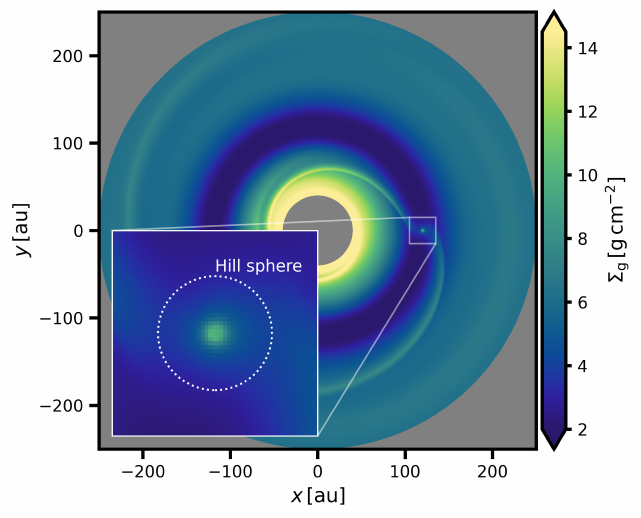}
    \includegraphics[width=0.98\linewidth]{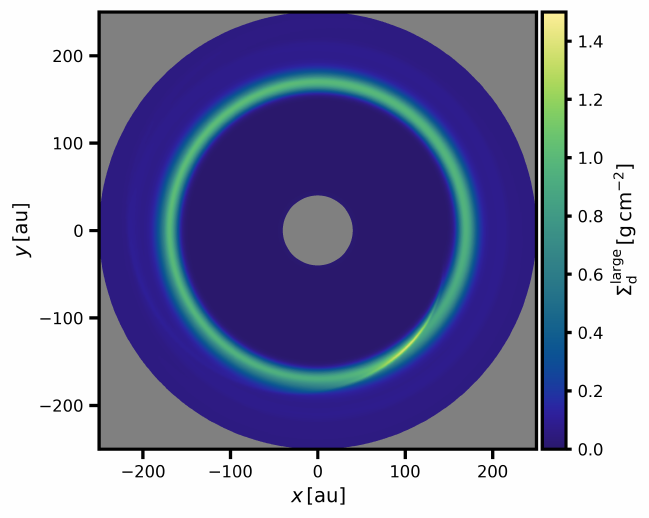}
    \caption{Surface density distribution of gas (\emph{top}) and large dust
    grains (\emph{bottom}) at the end of our nominal simulation. The width of the inset in
    the \emph{top panel} is 30 au.}
    \label{fig:M1_midplane}
\end{figure}

\begin{figure*}
    \centering
    \includegraphics[width=0.49\linewidth]{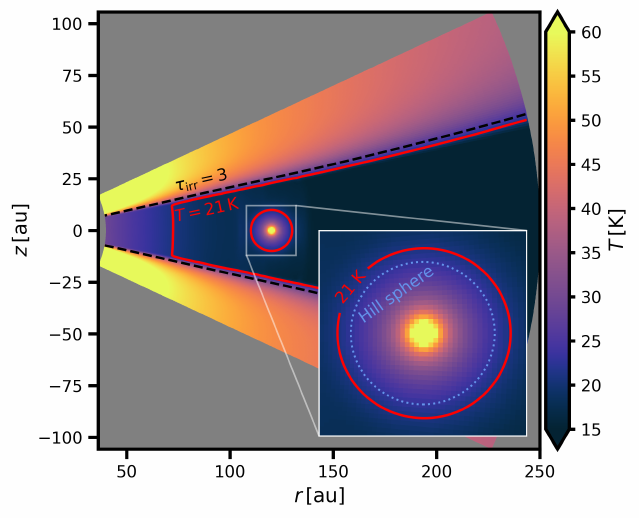}    \includegraphics[width=0.49\linewidth]{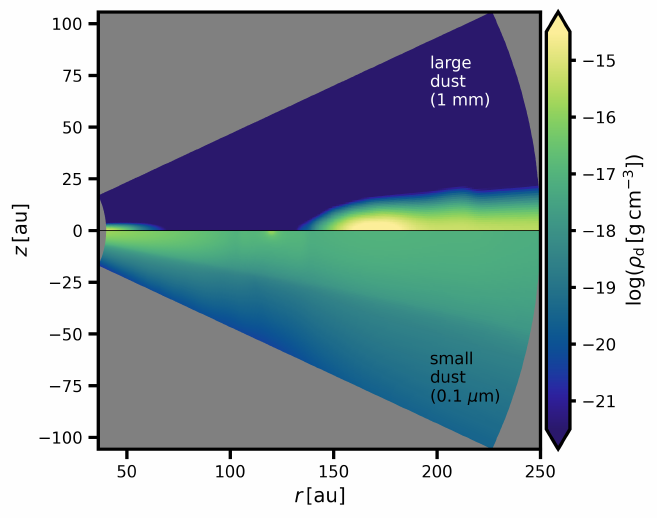}
    \caption{Meridional profile of the disk temperature (\emph{left}) and
    the volume density of small (\emph{right}, \emph{bottom half}) and large dust grains (\emph{right}, \emph{upper half}). The
    displayed plane coincides with the location of the planet.
    The temperature map also shows the isosurfaces at which $T=21\,\mathrm{K}$
    (solid red curve) and the optical depth to stellar irradiation becomes $\tau_{\mathrm{irr}}=3$ (dashed black curve). The width of the inset, in which we also mark the Hill sphere of the planet (dotted blue curve),  is $24\,\mathrm{au}$. The accreting planet
    heats its surroundings and maintains a bubble within which $T>21\,\mathrm{K}$ and the freeze-out of CO molecules is prevented, unlike in the majority of the disk interior.}
    \label{fig:M1_slice}
\end{figure*}

\subsection{Synthetic observations}
\label{sec:synth_obs}

We postprocessed the results of our hydrodynamic simulations
with the Monte Carlo radiative transfer code \textsc{Radmc-3D}\footnote{https://www.ita.uni-heidelberg.de/~dullemond/software/radmc-3d/} \citep{Dullemond_etal_2012ascl.soft02015D}
to produce synthetic images for CO emission lines and the dust continuum.
We studied the molecular isotopologs $^{12}$CO, $^{13}$CO, and C$^{18}$O and 
focused on their rotational transition $J=2$--$1$ with the rest frequencies
$\simeq$$230.538$, $220.399$, and $219.560\,\mathrm{GHz}$, respectively.

For our calculations with \textsc{Radmc-3D}, we adopted the same computational grid
as in our 3D hydrodynamic simulations. We input the density distributions of both small and large dust grains
and used their full frequency-dependent opacities computed with \textsc{optool} (Sect.~\ref{sec:opacity}). Since our hydrodynamic model is radiative, it provides a reasonable profile
of the disk temperature. Moreover, it contains the effects of compressional 
heating, viscous heating, and accretion luminosity.
Therefore, we did not recompute the temperature with \textsc{Radmc-3D}. 
In order to derive the number density of $^{12}$CO molecules, we assumed
an abundance of $10^{-4}$ relative to H$_{2}$. The isotopic ratios for the remaining molecules
were $[^{12}\mathrm{C}]/[^{13}\mathrm{C}] = 77$ and $[^{16}\mathrm{O}]/[^{18}\mathrm{O}] = 560$ \citep{Wilson_Rood_1994ARA&A..32..191W}.
The population levels were calculated in local thermodynamic equilibrium
using the molecular data from the LAMDA\footnote{https://home.strw.leidenuniv.nl/~moldata/} database
\citep{Schoier_etal_2005A&A...432..369S}.
We ignored the effects of microturbulence (which might affect the line profiles)
and photodissociation (which might affect the molecular abundances in the upper disk layers
that are exposed to high-energy radiation).
An important effect that we included, 
albeit using a primitive scaling,
was the freeze-out of CO molecules. In all locations in which the 
temperature was $T\leq21\,\mathrm{K}$ \citep[e.g.][]{Schwarz_etal_2016ApJ...823...91S,Pinte_etal_2018A&A...609A..47P},
we reduced the molecular abundance by a scaling factor
$f_{\mathrm{freeze}}=10^{-5}$ \citep[e.g.][]{Barraza-Alfaro_etal_2024A&A...683A..16B}

Unless stated otherwise, our synthetic images 
were generated assuming disk inclination 
$i=45^{\circ}$ and distance $d_{\mathrm{disk}}=100\,\mathrm{pc}$.
Scattering effects were neglected for simplicity.

\section{Nominal simulation} 
\label{sec:nominal}

\begin{figure*}
    \centering    
    \includegraphics[width=0.49\linewidth]{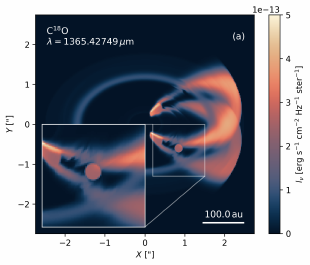}
    \includegraphics[width=0.49\linewidth]{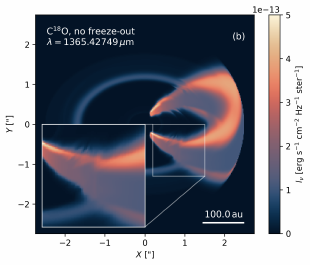}
    \includegraphics[width=0.49\linewidth]{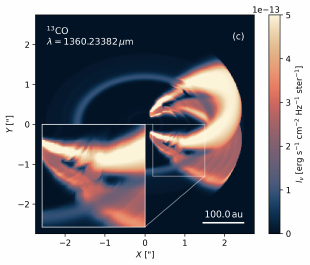}
    \includegraphics[width=0.49\linewidth]{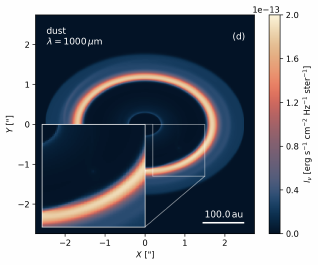}
    \caption{Synthetic \textsc{Radmc-3D} images based on our nominal simulation. The disk
    inclination is $45^{\circ}$, and the angular position of the planet is $45^{\circ}$ clockwise from the disk major axis.
    (a) Example channel map of C$^{18}$O. The inset
    shows the CO bubble in detail.
    (b) As in the first panel, but ignoring freeze-out. (c) Example channel map of $^{13}$CO. (d) Dust continuum emission at 1 mm.
    Each image also contains the wavelength at which it was taken. The width of the inset is $1.3$ arcsec.
    The color scale, which represents the emission intensity, is kept fixed between panels (a)--(c).}
    \label{fig:M1_images}
\end{figure*}

Our nominal simulation was based on parameters that are summarized in Table~\ref{tab:params_nominal}
and included a Jupiter-mass planet ($M_{\mathrm{p}}=1\,M_{\mathrm{Jup}}$) orbiting
at $r_{\mathrm{p}}=120\,\mathrm{au}$. At the given distance, our resolution led
to grid cells that were nearly cube-shaped, which helped to minimize numerical anisotropy,
especially for the radiation and dust diffusion. The Hill sphere of the planet was resolved
by 30 cells in each dimension. The considered mass-doubling time of the planet,
$t_{\mathrm{acc}}=0.1\,\mathrm{Myr}$, translates to the planetary luminosity $L_{\mathrm{p}}=2.8\times10^{-3}\,L_{\odot}$ or the effective surface temperature $T_{\mathrm{p,e}}\simeq4000\,\mathrm{K}$ (assuming the bulk density of the planet $\rho_{\mathrm{p}}=1\,\mathrm{g}\,\mathrm{cm}^{-3}$). Placing this temperature in the context of previous works dedicated
to studying the circumplanetary region at high resolution, we recall that \cite{Ayliffe_Bate_2009MNRAS.397..657A} found peak
temperatures $\sim$$4500\,\mathrm{K}$ and \cite{Szulagyi_2017ApJ...842..103S} considered
$\sim$$1000$--$10000\,\mathrm{K}$.

\subsection{Global disk structure}

Figure~\ref{fig:M1_profiles} shows 1D azimuthally averaged profiles
of several characteristic quantities at the end of the simulation stages introduced
in Sect.~\ref{sec:stages}.
The hydrodynamic relaxation stage covered $200$ orbits, and the main run including the planet spanned $800$ orbits. The planet mass
and luminosity were gradually increased over
the first $50$ orbits 
of the main run\footnote{The simulated time interval
with a full-grown planet (750 orbits) translates
to $\simeq$$1\,\mathrm{Myr}$. Full convergence of the gap profile was not reached during this time, but
the relative change in $\Sigma\gas(r)$ is only $5\%$ per 100 orbits at the end of the simulation.}.
Clearly, the proximity of curves corresponding to the hydrostatic and hydrodynamic
relaxation indicates that the former is already a decent estimate of the equilibrium
disk structure. The only noticeable difference arises for the surface density 
of large dust grains, $\Sigma\bg$, because $\Sigma\bg/\Sigma\gas$ was set constant during the hydrostatic relaxation. This does not necessarily imply a radially uniform mass flux of dust during hydrodynamic calculations, however,
because of radial variations
of the Stokes number and 
of the pressure gradient of
the nonisothermal
gaseous component
\citep[for analytical arguments, see appendix A of][]{Chrenko_etal_2024A&A...690A..41C}.

Based on the profile of the aspect ratio, it is important to assess (i) whether
the vertical span of the domain is larger than the typical height of the main
CO emitting surfaces
and (ii) how well we resolve the settled dusty layer of large grains.
At the planet location, the aspect ratio is $h_{\mathrm{p}}\simeq0.09$, implying a local pressure scale height $H_{\mathrm{p}}\simeq11\,\mathrm{au}$. 
Since the line-forming surfaces of CO are typically located at $\simeq$$2$--$4H$
\citep{Law_etal_2022ApJ...932..114L}, we are indeed able to capture them because
the vertical boundary of our domain reaches $\simeq$$5H$.
Concerning point (ii), the typical Stokes number at the planet location
is $\mathrm{St}_{\mathrm{p}}\simeq0.014$, which results in $H_{\mathrm{d},\mathrm{p}}\simeq0.19 H_{\mathrm{p}}$ (Eq.~\ref{eq:Hdust}).
One scale height of large grains is thus sampled by four grid cells. This resolution
is rather poor, but given the global nature of our simulation and its relatively
long time span, we consider it a reasonable compromise between accuracy and computational
demands\footnote{Our nominal simulation
required six days on 24 NVIDIA A100 GPUs.}.

Figure~\ref{fig:M1_profiles} also reveals that the planet carves a gas gap whose outer
edge acts as an efficient trap for inward-drifting large grains. These grains pile up at
$r\simeq170\,\mathrm{au}$ and are nearly absent from the inner disk. 
The global impact of the planet on the azimuthally averaged temperature profile is marginal (but see Sect.~\ref{sec:warm_bubble}), only the gap region
becomes slightly cooler on average. This is due to the excavation 
of the gap region, which reduces the grazing angle of the surface where 
most of the irradiating photons are absorbed; basically, 
the gap region is weakly shadowed by the inner gap edge.
It is important to point out here, however, that the planetary gap
is relatively shallow even though it is carved by a Jupiter-mass perturber because the planet only marginally
exceeds the local thermal mass $M_{\mathrm{th}}=h^{3}M_{\star}$ \citep{Goodman_Rafikov_2001ApJ...552..793G}. It has $M_{\mathrm{p}}/M_{\mathrm{th}}\simeq1.4$, as dictated by the large local scale height $h_{\mathrm{p}}$.

The distribution of $\Sigma\gas$ and $\Sigma\bg$ in the $r$-$\phi$ plane is shown 
in Fig.~\ref{fig:M1_midplane}. In addition to showing the structure
of the gas gap and the dust ring in detail, it also enables us to study the gas and dust concentration
within the Hill sphere of the planet. The local peak of the gas density is rather modest
(e.g., compared to what is typically seen in locally isothermal simulations) due to
the energy output of the planet, which adds to the pressure support. The large dust grains
are absent from the circumplanetary region owing to the dust trap at the outer gap edge and to 
the midplane outflows discussed
in Sect.~\ref{sec:outflows}.

\subsection{Warm CO bubble surrounding the planet}
\label{sec:warm_bubble}

Figure~\ref{fig:M1_slice} (left panel) shows the disk temperature in a meridional plane
passing through the position of the planet. The disk exhibits a layered structure that is
typical for stellar irradiated disks \citep[e.g.][]{Chiang_Goldreich_1997ApJ...490..368C,Bitsch_etal_2013A&A...549A.124B,Flock_etal_2013A&A...560A..43F}, with a colder
interior below a warmer atmosphere. The atmosphere is where the stellar
photons stream freely until they are gradually absorbed, as marked by the 
dashed black curve. which is the surface at which the radially integrated optical depth to 
stellar irradiation becomes $\tau_{\mathrm{irr}}=3$.

The most important feature in the temperature profile is the warm
bubble that surrounds the luminous planet. The extent of the isosurfaces (red curves) 
at which $T=21\,\mathrm{K}$, which is the CO freeze-out temperature we assumed \citep{Schwarz_etal_2016ApJ...823...91S,Pinte_etal_2018A&A...609A..47P}, 
clearly shows that the thermal feedback of the planet protects CO molecules
from freeze-out in a region that is slightly larger than the Hill sphere
(for our nominal set of parameters). When we convert
the gas density into the number density of various CO isotopologs (see Sect.~\ref{sec:synth_obs}), the number density within the warm bubble therefore remains
higher by the factor $1/f_{\mathrm{freeze}}$ than for the majority of the 
disk interior at $r\gtrsim70\,\mathrm{au}$.

The right panel of Fig.~\ref{fig:M1_slice} compares the vertical distribution of small and large dust grains and again shows the filtering of large grains
at the planet-induced pressure bump, as well as the difference in the vertical settling between
the two populations. The importance of the dust distributions for the 
temperature profile is such that the small grains determine the location of the $\tau_{\mathrm{irr}}=3$ surface, whereas the presence (or absence) of large grains
is significant for the local rate of radiative cooling, as we further explore in Appendix~\ref{sec:additional}.

\section{Observability} \label{sec:obs}

\subsection{Synthetic images}

Figure~\ref{fig:M1_images} provides an overview of synthetic images that we obtained
by post-processing the results of our nominal simulation with \textsc{Radmc-3D}. Panel (a) shows that
the warm bubble surrounding the accreting planet clearly stands out in C$^{18}$O emission and appears
as a low-intensity spot with a considerable solid angle. The edges of the dragonfly wings, the 
void in-between them, and the bubble itself directly trace the features of the temperature map shown in Fig.~\ref{fig:M1_slice}.

Panel (b) of Fig.~\ref{fig:M1_images}, 
in which we did not account for freeze-out when we converted the simulation data into the number density of C$^{18}$O,
reveals that all information about the circumplanetary region is lost without freeze-out. 
In this case, the edges of the dragonfly wings are connected by a wall of low-intensity emission that arises from the 
cold disk interior. The interior layers of CO also extinct the backside of the dragonfly wing, which only remains visible
as a narrow arc \citep[see also][]{Dullemond_etal_2020A&A...633A.137D}.

Panel (c) of Fig.~\ref{fig:M1_images} shows the emission of the $^{13}$CO isotopolog.
$^{13}$CO is generally more abundant than C$^{18}$O, and it therefore becomes optically thick already at higher elevations above the midplane.
For this reason, the dragonfly wings probe a slightly warmer layer in our simulation and appear to be brighter than C$^{18}$O (we used the same extent of the color scale in panels (a)--(c) to facilitate the comparison, but at the expense of saturating the $^{13}$CO image). The higher abundance of $^{13}$CO also partially obscures 
the bubble by the foreside surface layer
at the given geometry, suggesting that C$^{18}$O is more favorable
for distinguishing the bubble from the extended disk emission (which is confirmed in Sect.~\ref{sec:kinematics}). On its own, the bubble itself appears to be nearly identical to C$^{18}$O because the optical thickness of unity is reached already at the outskirts
of the bubble for all isotopologs. Therefore, the observed emission originates at gas temperatures that are only slightly
above $21\,\mathrm{K}$ (see Fig.~\ref{fig:M1_slice}, where the bubble is warmest in its center
and becomes cooler toward its edges). The bubble then
appears faint and exhibits a similar emission intensity for all isotopologs.
However, the bubble can become brighter when it is positioned at a channel edge
as we show in the following section.

\subsection{Synthetic ALMA data cubes and subtraction of the axial flow}
\label{sec:kinematics}

Actual observations are filtered by the instrument response and include noise. 
We neglected systematic errors due to sparse $uv$-plane sampling and assumed that these biases can be overcome with an adequate processing of ALMA high-fidelity imaging data. We also chose a beam matched to the Hill sphere size of a Jupiter-mass body at $\sim$$100$\,au. A larger beam would improve the signal-to-noise ratio (S/N) of the disk emission, but at the 
expense of diluting the planetary signal. The point-spread function (PSF) was thus taken to be a circular Gaussian 70\,mas in full width at half-maximum (FWHM). The radiative transfer images in native resolution were degraded with the  addition of thermal noise, corresponding to the target root mean square (rms) noise amplified by a factor close to $\sqrt{N}$, where $N$ is the number of pixels in the solid angle covered by the Gaussian beam. The noise amplification factor was calibrated on a blank image, and it thus ensured the target noise level in the smoothed data cube.

\begin{figure*}
    \centering
    \includegraphics[width=\linewidth]{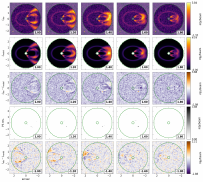}
    \caption{Analysis of synthetic ALMA data cubes.
    \emph{Top row:} Filtered sky images in C$^{18}$O(2-1) for selected channels. 
\emph{Second row:} Axially symmetric model data cube obtained with \textsc{disckin}.  \emph{Third row:} Residuals. \emph{Fourth row:} Point-source residuals after median filtering (see text), with the peaks from 1.0 to 1.6\,km\,s$^{-1}$ corresponding to the CO bubble
and reaching $9\sigma$ significance relative to the scatter in the residuals.
\emph{Bottom row:} Symmetric velocity channels relative to the systemic velocity. The data cubes have been averaged into coarse sky pixels, whose solid angle approximates that of the beam, as a means to speed up the \textsc{disckin} optimization.}
    \label{fig:channelmaps}
\end{figure*}

For example, in the case of C$^{18}$O(2-1), the ALMA sensitivity calculator \citep[][]{2019athb.rept.....R} yields an expected noise of 0.37\,mJy\,beam$^{-1}$ under ideal conditions in a 0.2\,km\,s$^{-1}$ bandwidth and with a 2\,d integration time, which we regard as a practical maximum on sensitivity. 
In this application of RADMC-3D, we directly sampled the data cube at intervals of 0.04\,km\,s$^{-1}$ and then binned the spectral direction into 0.2\,km\,s$^{-1}$ channels (i.e., we averaged five native radiative transfer channels). 
Selected synthetic channels covering the signal from circumplanetary material are shown in the top row of Fig.\,\ref{fig:channelmaps}.

\begin{figure}
    \centering
    \includegraphics[width=\columnwidth]{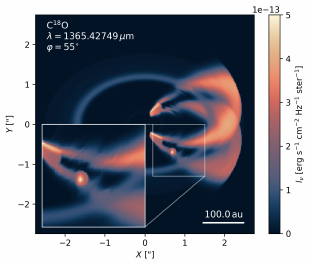}
    \caption{As panel (a) of Fig.~\ref{fig:M1_images}, but the underlying hydrodynamic
    model is rotated $10^{\circ}$ toward the minor axis of the sky projection ($\varphi=55^{\circ}$
 compared to $45^{\circ}$ in Fig.~\ref{fig:M1_images}).
    An animated version scanning a broader range of rotations is available \texttt{online}.}
    \label{fig:brightening}
\end{figure}

\begin{figure}
    \centering
    \includegraphics[width=0.77\columnwidth]{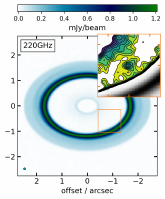}
    \caption{Synthetic ALMA observation of the 220\,GHz continuum in our nominal simulation with thermal  noise, but without $uv$-plane filtering. The inset is centered on the circumplanetary environment, which is barely recognizable (at 8$\sigma$, but confused with structures along the planetary wake). A beam ellipse is shown in the bottom left corner and also in the inset. The contour levels in the inset start at 3$\sigma$ and then rise by one $\sigma$ in alternating colors. The color stretch in the figure and the grayscale in the inset are linear and cover the full range of intensities.  The tick marks in the inset are placed at 0.3\arcsec intervals.}   \label{fig:synthetic_continuum}
\end{figure}

The warm CO bubble is readily identified  by visual inspection, for instance, at $v =1.2$~km\,s$^{-1}$. It is surrounded by the extended and structured disk emission, however, which would hamper its detection in automatized searches. 
The unresolved signal of the bubble can be made more conspicuous by subtracting the disk emission. The same tools that are used for the kinematic planet searches can be applied to build such a background model data cube, stemming from an axially averaged disk. We used the package \textsc{disckin} (Casassus et al. in prep), which extracts the disk rotation curve by assuming axially symmetric line formation surfaces in the local thermodynamic equilibrium and includes meridional flows. In the case at hand, for C$^{18}$O(2-1), these surfaces approximate the CO layers on either side of the frozen midplane beyond a minimum radius $R_{\rm freeze}$ (which is a free parameter),
but include the midplane layer interior to $R_{\rm freeze}$. The inference  of the nonparametric radial profiles follows from the package \textsc{ConeRot} \citep[][]{Casassus_Perez_2019ApJ...883L..41C, Casassus2021MNRAS.507.3789C}, but was derived from a least-squares fit to the entire data cube rather than the velocity centroid images. 
The structure and physical conditions in the underlying axially symmetric \textsc{disckin} model will be benchmarked against the input hydrodynamics in an upcoming article.

Based on the axially averaged channel maps, labeled $I_{\rm model}$ in Fig.\,\ref{fig:channelmaps}, we subtracted the extended disk emission and used the residuals to search for significant point sources (PSs). The systematic errors due to nonthermal residuals dominated, and we therefore applied a median filter to the residuals  $I_{\rm obs} - I_{\rm model}$ (third row in Fig.\,\ref{fig:channelmaps}) in order to detect a point-like signal. We used a square kernel of five times the beam width on each side and selected pixels that deviated by more than 3$\,\sigma$ from the  median-filtered image, where the standard deviation $\sigma$ was computed in each channel. The resulting PS residuals are shown in the fourth row of  Fig.\,\ref{fig:channelmaps}. The peak PS residual is at 9$\sigma$ at +1.6\,km\,s$^{-1}$, and the bubble is readily identified in that channel. The bubble is fainter in the other channels and reaches 6$\sigma$ at +1.2\,km\,s$^{-1}$ and +1.4\,km\,s$^{-1}$, where it can be confused with other $\lesssim$4$\sigma$ point-like structures from the disk.

It is interesting to note that the bubble is brighter at velocities corresponding to the edges of the channel maps. This is due to the sampling of the larger nonaxial velocity deviations closer to the accreting body, which stem from the hotter gas inside the bubble. By contrast, at disk velocities, the bubble covers a larger extent at lower brightness temperatures. The large solid angle should be favorable to its detection
(see panel (a) of Fig.\ref{fig:M1_images}). However, the 
disk is nearly Keplerian at the co-orbital radius and the bubble is confused with disk emission. The \textsc{disckin} kinematic model then adjusts the parameters to fit the bubble as part of the axially symmetric channel maps, which further decreases its brightness in the residuals. This rather surprising effect is due to the extension of the bubble at disk velocities, where it merges with the disk emission.

Figure~\ref{fig:brightening} demonstrates this brightening of the bubble at an
idealized resolution and without noise. To allow for a direct comparison with Fig.~\ref{fig:M1_images},
we used the same channel, but rotated the hydrodynamic model so that 
only a part of the flow within the bubble contributed to the emission.
Consequently, the bubble is divided in half and the emission probes closer
to its warmer central parts
that deviate more strongly
from the mean disk rotation.

Another interesting aspect of our analysis of the observable kinematics 
is that the emergent data cubes can be reproduced almost entirely with an axially symmetric flow. The residuals in the symmetric velocity channels in Fig.\,\ref{fig:channelmaps} (bottom row) barely skim the 3$\sigma$ level even in this extremely deep and idealized observation. In other words, the residuals in a more standard observation, with a few hours of integration, would be entirely thermal. Thus, the nonaxisymmetric kinematic perturbations
that are due to the gravitational perturbations induced by a body that marginally exceeds the local
thermal mass are not detectable. We note, however, that the modulation of the azimuthal rotation curve due to  variable hydrostatic support across the gap, as described by \citet{Teague2018ApJ...860L..12T}, is indeed readily detected in the axially symmetric kinematics.

\begin{figure*}
    \centering
    \includegraphics[width=0.49\linewidth]{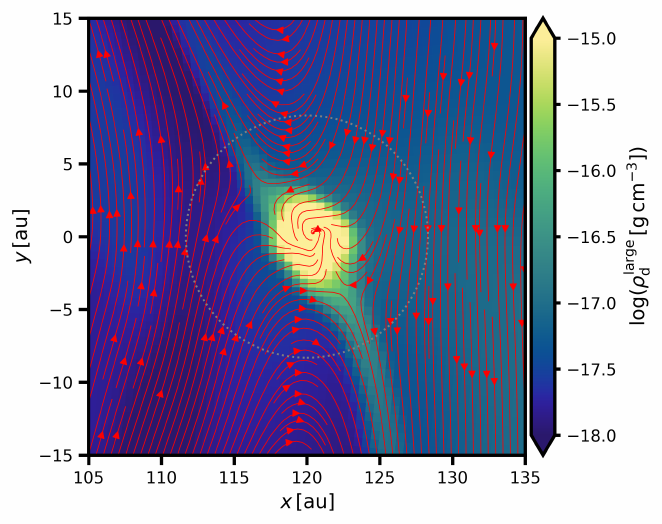}
    \includegraphics[width=0.49\linewidth]{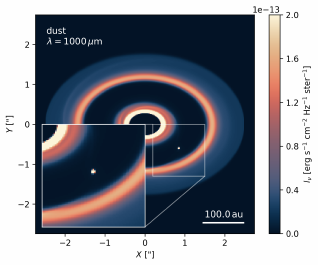}
    \caption{\emph{Left:} Logarithm of the midplane density of large dust $\rho\bg$
    in the vicinity of the planet at $t=375$ orbits. The 
    dust streamlines (red arrows) and the Hill sphere of the planet
    (dotted gray circle) are overlaid. Dust leakage is apparent
    along the downstream horseshoe flow.
    \emph{Right:} Continuum emission of large dust grains
    at $t=375$ orbits, with a hot spot
    arising from the circumplanetary region.
    The extent of the color scale is the same as in panel (d) of Fig.~\ref{fig:M1_images}
    to enable a direct comparison.
    }
    \label{fig:dustplume}
\end{figure*}

\begin{figure}
    \centering
    \includegraphics[width=\columnwidth]{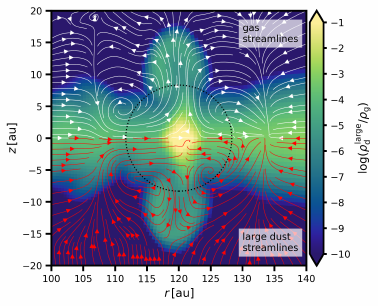}
    \caption{Logarithm of the vertical dust-to-gas ratio $\rho\bg/\rho\gas$ in the vicinity of the planet at $t=375$ orbits (same $t$ as in Fig.~\ref{fig:dustplume}).
    The streamlines of the gas and dust meridional flows (white and red arrows, respectively) are overlaid.
    The Hill radius is marked with a dotted black circle. The formation of two 
    plumes associated with polar outflows is apparent; nevertheless, the dust fraction contained
    in the plumes is small.}
    \label{fig:dustplume_vert}
\end{figure}

\subsection{Continuum emission and the depletion of circumplanetary dust}
\label{sec:outflows}

In addition to the molecular emission, the channel maps contain a ring
that corresponds to the underlying dust continuum emission of
large grains that accumulate at the outer edge of the planet-induced gap.
To examine the dust emission in detail, we show a synthetic image at $\lambda=1\,\mathrm{mm}$
in panel (d) of Fig.~\ref{fig:M1_images}. In addition to the ring itself, there is an emission void interior to it
and a faint skirt exterior to it. The latter is due to dust grains that we continued to reintroduce at the 
outer edge of the disk during the entire simulation. Importantly, the warm circumplanetary region 
is not detected in the thermal dust emission
because there is a paucity of large dust grains in the circumplanetary region.

To further confirm the emission topology of Fig.~\ref{fig:M1_images} (d), we
generated a synthetic ALMA continuum observation. We followed the same procedure as in Sect.\,\ref{sec:kinematics}, but used a 7.5\,GHz bandwidth to reach a thermal noise of 1.7\,$\mu$Jy\,beam$^{-1}$ in the continuum underlying C$^{18}$O(2-1) (220\,GHz), following a 2\,d integration in the best conditions. 
Fig.~\ref{fig:synthetic_continuum} shows that even in this ideal observation with perfect calibration and no synthesis imaging artifacts\footnote{We stress that the required signal-to-noise ratio $\sim$700 have only been achieved in compact array configurations and after self-calibration. Resolved long-baseline observations with ALMA are limited by the imperfect visibility calibration (e.g., due to the phase noise) and seldom reach beyond $\mathrm{S/N}\sim100$. However, Dom\'{i}nguez-Jamett et al. (in prep) have recently reached $\mathrm{S/N}\sim350$ observing PDS\,70 with a 25\,h integration in Band\,7 and a similar beam as assumed here.}, the environment surrounding the protoplanet would be barely detected because only the planetary wake would be picked up as a noisy filament \citep[resembling the noisy filament in HD\,135344B;][]{Casassus2021MNRAS.507.3789C}.
Trials at the lower disk inclination of $i=30\,$deg showed that the protoplanet would be entirely undetected in the same setup. 
The fact that the circumplanetary environment of the marginally superthermal
planet resulting from our gas-dust hydrodynamic simulation would remain undetected even in deep ALMA observations (except barely in the ideal 2\,d integration considered here)
is in contrast with several previous simulation-based predictions
(see Sect.\,\ref{sec:intro}) and is fully consistent
with the paucity of CPD detections.

To understand why the large dust grains are missing,
we examined the evolution of the circumplanetary
dust content and found that it decays in time.
Fig.~\ref{fig:dustplume} (left panel) shows the 
circumplanetary density distribution of large dust 
at an earlier simulation time, $t=375$ orbits. 
Apparently, the concentration of large dust grains 
surrounding the planet is
relatively large early on: The planet is able 
to accumulate these grains before the gap region
is fully cleared of them and their radial flux is diminished.
However, Fig.~\ref{fig:dustplume} reveals that 
dust grains can leak from the central
overdensity that surrounds the planet by entering the horseshoe
flow as it U-turns deep within the Hill sphere. An increased 
dust density is evident mainly along the downstream horseshoe flow
that trails the planet (in the $x>120$ and $y<0\,\mathrm{au}$ quadrant). 
To confirm that this outflow plays a decisive role in the circumplanetary
dust depletion, we performed an extended analysis of the dust mass flux
in Appendix~\ref{sec:dust_depletion}.

The temporal evolution of the circumplanetary dust content
is reflected in the dust continuum emission.
The second panel of Fig.~\ref{fig:dustplume} shows that
the warmed-up circumplanetary dust manifests itself as a hot spot
at $t=375$ orbits. After approximately $t\simeq450$ orbits,
however, the dust depletion via horseshoe flows
leads to a disappearance of the hot spot, and the 
thermal continuum image becomes more similar to panel (d)
of Fig.~\ref{fig:M1_images}.

The dust flow in the close proximity
of the planet can still effectively exchange material with the horseshoe 
flow because no full-fledged CPD is formed
for the given combination of planet mass, disk conditions,
and accretion heating. Additional differences between the circumplanetary
environment in our nominal model and classical CPDs are highlighted 
in Fig.~\ref{fig:dustplume_vert}, where we identify a two-sided polar 
outflow in the meridional plane. The gas and dust both follow this outflowing 
pattern, as reflected by the formation of two dusty plumes that rise 
above the orbital plane. The polar outflow can only entrain
dust grains at heights that are already relatively dust poor, however,
and thus, the plumes themselves do not substantially
contribute to the depletion of the circumplanetary dust content (the dust-to-gas
ratio in the plumes is lower by roughly four orders of magnitude than the dust accumulation surrounding the planet).

Last but not least, it is worth noting that 
even in the absence of accretion heating, the 
Jupiter-mass planet would not be capable of forming
a full-fledged CPD at $r_{\mathrm{p}}=120\,\mathrm{au}$
within our disk model. The level of vertical
gas compression inside the Hill sphere would instead remain only moderate and
transition between a quasi-spherical envelope
and a highly compressed local disk \citep[see also][]{Sagynbaeva_etal_2024arXiv241014896S}.
This is further documented in Appendix~\ref{sec:scold}.

\section{Discussion and conclusions}
\label{sec:conclusions}

We studied the influence of a luminous accreting protoplanet
on the CO emission arising from a protoplanetary disk.
We developed an advanced 3D radiation hydrodynamics model
involving both gas and dust in a two-fluid approximation. 
We assumed that small (submicron-sized) dust grains passively
trace the gas distribution, and we evolved large (millimeter-sized) dust grains directly,
using both grain populations to derive optical depths to 
stellar irradiation and thermal radiation diffusion.
As our nominal case, we chose a Jupiter-mass planet
with an effective temperature $T_{\mathrm{p,e}}=4000\,\mathrm{K}$ that
orbited at $r_{\mathrm{p}}=120\,\mathrm{au}$.
In our disk model, the ratio 
of the planet mass to the local thermal mass was about unity.
The orbital radius fell outside the CO snowline, into 
a cold layer in which CO molecules freeze out and are only present in low abundances.

The accretion luminosity of the planet heats its vicinity, where it maintains a bubble of warm gas with temperatures above the freeze-out threshold \citep[we assumed $21\,\mathrm{K}$;][]{Schwarz_etal_2016ApJ...823...91S,Pinte_etal_2018A&A...609A..47P}.
Consequently, the abundance of gas-phase CO molecules within
the bubble remains higher than in the broad cold background.
By inspecting synthetic CO channel maps, we found that 
the bubble can be readily detected by subtracting 
an axially symmetric kinematic disk profile 
and applying median filtering. The peak residual
identifies the bubble as a point source located within
the emission void  between the fore- and backside line-forming surfaces.
This requires 
that the disk is moderately inclined ($i\approx45^{\circ}$),
the planet is not too close to the minor axis of the disk,
and the extended disk emission does not blend
with (or extincts) the bubble. The latter constraint favors
rarer isotopologs such as C$^{18}$O, which can become optically thin in the planetary gap region.

Since the bubble itself is optically thick, its emission
is faint because it originates from the outer shells
that are only marginally warmer than the freeze-out temperature. For the same reason,
its emission intensity is nearly independent of the generating isotopolog.

In addition to the nominal simulation, we presented a modest suite
of additional runs in Appendix~\ref{sec:additional}. They demonstrate that (i) the bubble shrinks when the accretion
luminosity decreases, which can be counteracted by modifying
the opacity law (e.g.. by assuming a higher-than-nominal constant opacity); (ii)
when the planet is shifted just outside the CO snowline,
the bubble can locally perturb the shape of the snowline itself;
(iii) the bubble can form around subthermal-mass planets as well,
but it is small and hardly detectable  then (for a single value
of the mass-doubling time that we tested).

Since our model tracks the evolution of mm-sized dust grains,
it also provides predictions for the thermal continuum emission.
In this regard, we captured the formation of a
dusty ring in a pressure bump at the outer edge of the planetary gap, as well as the depletion of dust from the circumplanetary region by the downstream horseshoe flow.
Therefore, the detectability of circumplanetary dust in the continuum depends 
on the level of depletion, exhibiting a hot spot during the first $\simeq$$450$ orbits, but showing no substantial signal at later times.
This dust depletion might explain
why circumplanetary dusty regions are observationally elusive.

We emphasize that in addition to the continuum nondetection,
the marginally superthermal planet does not produce detectable velocity
kinks. The resurgence of gaseous CO driven by thermal accretion feedback
thus offers promising prospects for automated protoplanet detections that would otherwise seem impossible.
Additionally, detections of CO bubbles would also provide crucial insights into the chemical processes that occur within circumplanetary environments, bridging the early stages of formation with the final properties of (exo)planets \citep[e.g.][]{Oberg_etal_2011ApJ...743L..16O}.
Our study adds to previous efforts to link the accretion luminosity
of forming planets to the observable chemistry \citep{Cleeves_etal_2015ApJ...807....2C,Jiang_etal_2023A&A...678A..33J}
or to disk kinematics \citep{Muley_etal_2024A&A...687A.213M}.

Although our work represents an important step toward
an improved realism of simulated molecular and continuum emission, it includes several caveats.
Regarding the molecular emission, no chemical module was involved in our calculations, and thus, the molecular abundances and the freeze-out effect depend on our choice of
scaling factors when the simulated gas density is converted
into the number density of the respective isotopologs.
Moreover, we assumed instantaneous freeze-out and sublimation,
based on the local temperature. 
In future work, we would like to understand whether
the disk shear can distort the bubble shape or even spread
it over a larger area \citep[see][]{Jiang_etal_2023A&A...678A..33J}.
Finally, the accretion luminosity, which is of critical
importance for the formation of the bubble, is 
parametric and was not obtained in a self-consistent manner.

\begin{acknowledgements}
We wish to thank the anonymous referee
whose constructive comments allowed us to improve
the manuscript.
This work was supported by the Czech
Science Foundation (grant 25-16507S), the Charles University
Research Centre program (No. UNCE/24/SCI/005), and the
Ministry of Education, Youth and Sports of the Czech Republic
through the e-INFRA CZ (ID:90254).  S.C. acknowledges
support from Agencia Nacional de Investigaci\'on y Desarrollo de Chile
(ANID) given by FONDECYT Regular grants 1211496 and  ANID project Data
Observatory Foundation DO210001. The authors are grateful to Marcelo Barraza-Alfaro,
Sebastiaan Krijt, and Mario Flock for useful and motivating discussions.
We acknowledge the use of \textsc{Python} libraries
\textsc{matplotlib} \citep{Hunter_2007_Matplotlib}, \textsc{numpy} \citep{Harris_etal_2020_Numpy},
and \textsc{lic} (\url{https://gitlab.com/szs/lic}).
\end{acknowledgements}

%
%

\bibliographystyle{aa}
\bibliography{references}

\begin{thebibliography}{98}
\expandafter\ifx\csname natexlab\endcsname\relax\def\natexlab#1{#1}\fi

\bibitem[{{Andrews}(2020)}]{Andrews_2020ARA&A..58..483A}
{Andrews}, S.~M. 2020, \araa, 58, 483

\bibitem[{{Ayliffe} \& {Bate}(2009)}]{Ayliffe_Bate_2009MNRAS.397..657A}
{Ayliffe}, B.~A. \& {Bate}, M.~R. 2009, \mnras, 397, 657

\bibitem[{{Bae} {et~al.}(2023){Bae}, {Isella}, {Zhu}, {Martin}, {Okuzumi}, \& {Suriano}}]{Bae_etal_2023ASPC..534..423B}
{Bae}, J., {Isella}, A., {Zhu}, Z., {et~al.} 2023, in Astronomical Society of the Pacific Conference Series, Vol. 534, Protostars and Planets VII, ed. S.~{Inutsuka}, Y.~{Aikawa}, T.~{Muto}, K.~{Tomida}, \& M.~{Tamura}, 423

\bibitem[{{Bae} {et~al.}(2022){Bae}, {Teague}, {Andrews}, {Benisty}, {Facchini}, {Galloway-Sprietsma}, {Loomis}, {Aikawa}, {Alarc{\'o}n}, {Bergin}, {Bergner}, {Booth}, {Cataldi}, {Cleeves}, {Czekala}, {Guzm{\'a}n}, {Huang}, {Ilee}, {Kurtovic}, {Law}, {Le Gal}, {Liu}, {Long}, {M{\'e}nard}, {{\"O}berg}, {P{\'e}rez}, {Qi}, {Schwarz}, {Sierra}, {Walsh}, {Wilner}, \& {Zhang}}]{Bae_etal_2022ApJ...934L..20B}
{Bae}, J., {Teague}, R., {Andrews}, S.~M., {et~al.} 2022, \apjl, 934, L20

\bibitem[{{Barraza-Alfaro} {et~al.}(2024){Barraza-Alfaro}, {Flock}, \& {Henning}}]{Barraza-Alfaro_etal_2024A&A...683A..16B}
{Barraza-Alfaro}, M., {Flock}, M., \& {Henning}, T. 2024, \aap, 683, A16

\bibitem[{{Benisty} {et~al.}(2021){Benisty}, {Bae}, {Facchini}, {Keppler}, {Teague}, {Isella}, {Kurtovic}, {P{\'e}rez}, {Sierra}, {Andrews}, {Carpenter}, {Czekala}, {Dominik}, {Henning}, {Menard}, {Pinilla}, \& {Zurlo}}]{Benisty_etal_2021ApJ...916L...2B}
{Benisty}, M., {Bae}, J., {Facchini}, S., {et~al.} 2021, \apjl, 916, L2

\bibitem[{{Benisty} {et~al.}(2023){Benisty}, {Dominik}, {Follette}, {Garufi}, {Ginski}, {Hashimoto}, {Keppler}, {Kley}, \& {Monnier}}]{Benisty_etal_2023ASPC..534..605B}
{Benisty}, M., {Dominik}, C., {Follette}, K., {et~al.} 2023, in Astronomical Society of the Pacific Conference Series, Vol. 534, Protostars and Planets VII, ed. S.~{Inutsuka}, Y.~{Aikawa}, T.~{Muto}, K.~{Tomida}, \& M.~{Tamura}, 605

\bibitem[{{Ben{\'\i}tez-Llambay} {et~al.}(2019){Ben{\'\i}tez-Llambay}, {Krapp}, \& {Pessah}}]{Benitez-Llambay_2019ApJS..241...25B}
{Ben{\'\i}tez-Llambay}, P., {Krapp}, L., \& {Pessah}, M.~E. 2019, \apjs, 241, 25

\bibitem[{{Ben{\'{\i}}tez-Llambay} {et~al.}(2015){Ben{\'{\i}}tez-Llambay}, {Masset}, {Koenigsberger}, \& {Szul{\'a}gyi}}]{Benitez-Llambay_etal_2015Natur.520...63B}
{Ben{\'{\i}}tez-Llambay}, P., {Masset}, F., {Koenigsberger}, G., \& {Szul{\'a}gyi}, J. 2015, \nat, 520, 63

\bibitem[{{Ben{\'{\i}}tez-Llambay} \& {Masset}(2016)}]{Benitez-Llambay_Masset_2016ApJS..223...11B}
{Ben{\'{\i}}tez-Llambay}, P. \& {Masset}, F.~S. 2016, \apjs, 223, 11

\bibitem[{{Birnstiel} {et~al.}(2018){Birnstiel}, {Dullemond}, {Zhu}, {Andrews}, {Bai}, {Wilner}, {Carpenter}, {Huang}, {Isella}, {Benisty}, {P{\'e}rez}, \& {Zhang}}]{Birnstiel_etal_2018ApJ...869L..45B}
{Birnstiel}, T., {Dullemond}, C.~P., {Zhu}, Z., {et~al.} 2018, \apjl, 869, L45

\bibitem[{{Bitsch} {et~al.}(2013){Bitsch}, {Crida}, {Morbidelli}, {Kley}, \& {Dobbs-Dixon}}]{Bitsch_etal_2013A&A...549A.124B}
{Bitsch}, B., {Crida}, A., {Morbidelli}, A., {Kley}, W., \& {Dobbs-Dixon}, I. 2013, \aap, 549, A124

\bibitem[{{Bitsch} {et~al.}(2018){Bitsch}, {Morbidelli}, {Johansen}, {Lega}, {Lambrechts}, \& {Crida}}]{Bitsch_etal_2018A&A...612A..30B}
{Bitsch}, B., {Morbidelli}, A., {Johansen}, A., {et~al.} 2018, \aap, 612, A30

\bibitem[{{Casassus} \& {C{\'a}rcamo}(2022)}]{Casassus_Carcamo_2022MNRAS.513.5790C}
{Casassus}, S. \& {C{\'a}rcamo}, M. 2022, \mnras, 513, 5790

\bibitem[{{Casassus} {et~al.}(2021){Casassus}, {Christiaens}, {C{\'a}rcamo}, {P{\'e}rez}, {Weber}, {Ercolano}, {van der Marel}, {Pinte}, {Dong}, {Baruteau}, {Cieza}, {van Dishoeck}, {Jordan}, {Price}, {Absil}, {Arce-Tord}, {Faramaz}, {Flores}, \& {Reggiani}}]{Casassus2021MNRAS.507.3789C}
{Casassus}, S., {Christiaens}, V., {C{\'a}rcamo}, M., {et~al.} 2021, \mnras, 507, 3789

\bibitem[{{Casassus} \& {P{\'e}rez}(2019)}]{Casassus_Perez_2019ApJ...883L..41C}
{Casassus}, S. \& {P{\'e}rez}, S. 2019, \apjl, 883, L41

\bibitem[{{Chen} \& {Dong}(2024)}]{Chen_Dong_2024ApJ...976...49C}
{Chen}, K. \& {Dong}, R. 2024, \apj, 976, 49

\bibitem[{{Chiang} \& {Goldreich}(1997)}]{Chiang_Goldreich_1997ApJ...490..368C}
{Chiang}, E.~I. \& {Goldreich}, P. 1997, \apj, 490, 368

\bibitem[{{Chrenko} \& {Chametla}(2023)}]{Chrenko_Chametla_2023MNRAS.524.2705C}
{Chrenko}, O. \& {Chametla}, R.~O. 2023, \mnras, 524, 2705

\bibitem[{{Chrenko} {et~al.}(2024{\natexlab{a}}){Chrenko}, {Chametla}, {Masset}, {Baruteau}, \& {Bro{\v{z}}}}]{Chrenko_etal_2024A&A...690A..41C}
{Chrenko}, O., {Chametla}, R.~O., {Masset}, F.~S., {Baruteau}, C., \& {Bro{\v{z}}}, M. 2024{\natexlab{a}}, \aap, 690, A41

\bibitem[{{Chrenko} {et~al.}(2024{\natexlab{b}}){Chrenko}, {Flock}, {Ueda}, {M{\'e}rand}, {Benisty}, \& {Chametla}}]{Chrenko_etal_2024AJ....167..124C}
{Chrenko}, O., {Flock}, M., {Ueda}, T., {et~al.} 2024{\natexlab{b}}, \aj, 167, 124

\bibitem[{{Chrenko} \& {Lambrechts}(2019)}]{Chrenko_Lambrechts_2019}
{Chrenko}, O. \& {Lambrechts}, M. 2019, \aap, 626, A109

\bibitem[{{Chrenko} \& {Nesvorn{\'y}}(2020)}]{Chrenko_Nesvorny_2020A&A...642A.219C}
{Chrenko}, O. \& {Nesvorn{\'y}}, D. 2020, \aap, 642, A219

\bibitem[{{Christiaens} {et~al.}(2024){Christiaens}, {Samland}, {Henning}, {Portilla-Revelo}, {Perotti}, {Matthews}, {Absil}, {Decin}, {Kamp}, {Boccaletti}, {Tabone}, {Marleau}, {van Dishoeck}, {G{\"u}del}, {Lagage}, {Barrado}, {Caratti o Garatti}, {Glauser}, {Olofsson}, {Ray}, {Scheithauer}, {Vandenbussche}, {Waters}, {Arabhavi}, {Grant}, {Jang}, {Kanwar}, {Schreiber}, {Schwarz}, {Temmink}, \& {{\"O}stlin}}]{Christiaens_etal_2024A&A...685L...1C}
{Christiaens}, V., {Samland}, M., {Henning}, T., {et~al.} 2024, \aap, 685, L1

\bibitem[{{Cleeves} {et~al.}(2015){Cleeves}, {Bergin}, \& {Harries}}]{Cleeves_etal_2015ApJ...807....2C}
{Cleeves}, L.~I., {Bergin}, E.~A., \& {Harries}, T.~J. 2015, \apj, 807, 2

\bibitem[{{Commer{\c c}on} {et~al.}(2011){Commer{\c c}on}, {Teyssier}, {Audit}, {Hennebelle}, \& {Chabrier}}]{Commercon_etal_2011A&A...529A..35C}
{Commer{\c c}on}, B., {Teyssier}, R., {Audit}, E., {Hennebelle}, P., \& {Chabrier}, G. 2011, \aap, 529, A35

\bibitem[{{Currie} {et~al.}(2022){Currie}, {Lawson}, {Schneider}, {Lyra}, {Wisniewski}, {Grady}, {Guyon}, {Tamura}, {Kotani}, {Kawahara}, {Brandt}, {Uyama}, {Muto}, {Dong}, {Kudo}, {Hashimoto}, {Fukagawa}, {Wagner}, {Lozi}, {Chilcote}, {Tobin}, {Groff}, {Ward-Duong}, {Januszewski}, {Norris}, {Tuthill}, {van der Marel}, {Sitko}, {Deo}, {Vievard}, {Jovanovic}, {Martinache}, \& {Skaf}}]{Currie_etal_2022NatAs...6..751C}
{Currie}, T., {Lawson}, K., {Schneider}, G., {et~al.} 2022, Nature Astronomy, 6, 751

\bibitem[{{Cuzzi} {et~al.}(1993){Cuzzi}, {Dobrovolskis}, \& {Champney}}]{Cuzzi_etal_1993Icar..106..102C}
{Cuzzi}, J.~N., {Dobrovolskis}, A.~R., \& {Champney}, J.~M. 1993, \icarus, 106, 102

\bibitem[{{de Val-Borro} {et~al.}(2006){de Val-Borro}, {Edgar}, {Artymowicz}, {Ciecielag}, {Cresswell}, {D'Angelo}, {Delgado-Donate}, {Dirksen}, {Fromang}, {Gawryszczak}, {Klahr}, {Kley}, {Lyra}, {Masset}, {Mellema}, {Nelson}, {Paardekooper}, {Peplinski}, {Pierens}, {Plewa}, {Rice}, {Sch{\"a}fer}, \& {Speith}}]{deValBorro_etal_2006MNRAS.370..529D}
{de Val-Borro}, M., {Edgar}, R.~G., {Artymowicz}, P., {et~al.} 2006, \mnras, 370, 529

\bibitem[{{Dipierro} {et~al.}(2016){Dipierro}, {Laibe}, {Price}, \& {Lodato}}]{Dipierro_etal_2016MNRAS.459L...1D}
{Dipierro}, G., {Laibe}, G., {Price}, D.~J., \& {Lodato}, G. 2016, \mnras, 459, L1

\bibitem[{{Dominik} {et~al.}(2021){Dominik}, {Min}, \& {Tazaki}}]{Dominik_OPTOOL_2021ascl.soft04010D}
{Dominik}, C., {Min}, M., \& {Tazaki}, R. 2021, Astrophysics Source Code Library, ascl:2104.010

\bibitem[{{Dubrulle} {et~al.}(1995){Dubrulle}, {Morfill}, \& {Sterzik}}]{Dubrulle_etal_1995Icar..114..237D}
{Dubrulle}, B., {Morfill}, G., \& {Sterzik}, M. 1995, \icarus, 114, 237

\bibitem[{{Dullemond} {et~al.}(2018){Dullemond}, {Birnstiel}, {Huang}, {Kurtovic}, {Andrews}, {Guzm{\'a}n}, {P{\'e}rez}, {Isella}, {Zhu}, \& {Benisty}}]{Dullemond_etal_2018ApJ...869L..46D}
{Dullemond}, C.~P., {Birnstiel}, T., {Huang}, J., {et~al.} 2018, \apj, 869, L46

\bibitem[{{Dullemond} {et~al.}(2020){Dullemond}, {Isella}, {Andrews}, {Skobleva}, \& {Dzyurkevich}}]{Dullemond_etal_2020A&A...633A.137D}
{Dullemond}, C.~P., {Isella}, A., {Andrews}, S.~M., {Skobleva}, I., \& {Dzyurkevich}, N. 2020, \aap, 633, A137

\bibitem[{{Dullemond} {et~al.}(2012){Dullemond}, {Juhasz}, {Pohl}, {Sereshti}, {Shetty}, {Peters}, {Commercon}, \& {Flock}}]{Dullemond_etal_2012ascl.soft02015D}
{Dullemond}, C.~P., {Juhasz}, A., {Pohl}, A., {et~al.} 2012, {RADMC-3D: A multi-purpose radiative transfer tool}

\bibitem[{{Fedele} {et~al.}(2023){Fedele}, {Bollati}, \& {Lodato}}]{Fedele_etal_2023A&A...672A.125F}
{Fedele}, D., {Bollati}, F., \& {Lodato}, G. 2023, \aap, 672, A125

\bibitem[{{Flock} {et~al.}(2013){Flock}, {Fromang}, {Gonz{\'a}lez}, \& {Commer{\c c}on}}]{Flock_etal_2013A&A...560A..43F}
{Flock}, M., {Fromang}, S., {Gonz{\'a}lez}, M., \& {Commer{\c c}on}, B. 2013, \aap, 560, A43

\bibitem[{{Flock} {et~al.}(2016){Flock}, {Fromang}, {Turner}, \& {Benisty}}]{Flock_etal_2016ApJ...827..144F}
{Flock}, M., {Fromang}, S., {Turner}, N.~J., \& {Benisty}, M. 2016, \apj, 827, 144

\bibitem[{{Goodman} \& {Rafikov}(2001)}]{Goodman_Rafikov_2001ApJ...552..793G}
{Goodman}, J. \& {Rafikov}, R.~R. 2001, \apj, 552, 793

\bibitem[{{Haffert} {et~al.}(2019){Haffert}, {Bohn}, {de Boer}, {Snellen}, {Brinchmann}, {Girard}, {Keller}, \& {Bacon}}]{Haffert_etal_2019NatAs...3..749H}
{Haffert}, S.~Y., {Bohn}, A.~J., {de Boer}, J., {et~al.} 2019, Nature Astronomy, 3, 749

\bibitem[{{Hammond} {et~al.}(2023){Hammond}, {Christiaens}, {Price}, {Toci}, {Pinte}, {Juillard}, \& {Garg}}]{Hammond_etal_2023MNRAS.522L..51H}
{Hammond}, I., {Christiaens}, V., {Price}, D.~J., {et~al.} 2023, \mnras, 522, L51

\bibitem[{Harris {et~al.}(2020)Harris, Millman, van~der Walt, Gommers, Virtanen, Cournapeau, Wieser, Taylor, Berg, Smith, Kern, Picus, Hoyer, van Kerkwijk, Brett, Haldane, del R{\'{i}}o, Wiebe, Peterson, G{\'{e}}rard-Marchant, Sheppard, Reddy, Weckesser, Abbasi, Gohlke, \& Oliphant}]{Harris_etal_2020_Numpy}
Harris, C.~R., Millman, K.~J., van~der Walt, S.~J., {et~al.} 2020, Nature, 585, 357

\bibitem[{Hunter(2007)}]{Hunter_2007_Matplotlib}
Hunter, J.~D. 2007, Computing in Science \& Engineering, 9, 90

\bibitem[{{Isella} {et~al.}(2019){Isella}, {Benisty}, {Teague}, {Bae}, {Keppler}, {Facchini}, \& {P{\'e}rez}}]{Isella_etal_2019ApJ...879L..25I}
{Isella}, A., {Benisty}, M., {Teague}, R., {et~al.} 2019, \apjl, 879, L25

\bibitem[{{Jiang} \& {Ormel}(2023)}]{Jiang_Ormel_2023MNRAS.518.3877J}
{Jiang}, H. \& {Ormel}, C.~W. 2023, \mnras, 518, 3877

\bibitem[{{Jiang} {et~al.}(2023){Jiang}, {Wang}, {Ormel}, {Krijt}, \& {Dong}}]{Jiang_etal_2023A&A...678A..33J}
{Jiang}, H., {Wang}, Y., {Ormel}, C.~W., {Krijt}, S., \& {Dong}, R. 2023, \aap, 678, A33

\bibitem[{{Keppler} {et~al.}(2018){Keppler}, {Benisty}, {M{\"u}ller}, {Henning}, {van Boekel}, {Cantalloube}, {Ginski}, {van Holstein}, {Maire}, {Pohl}, {Samland}, {Avenhaus}, {Baudino}, {Boccaletti}, {de Boer}, {Bonnefoy}, {Chauvin}, {Desidera}, {Langlois}, {Lazzoni}, {Marleau}, {Mordasini}, {Pawellek}, {Stolker}, {Vigan}, {Zurlo}, {Birnstiel}, {Brandner}, {Feldt}, {Flock}, {Girard}, {Gratton}, {Hagelberg}, {Isella}, {Janson}, {Juhasz}, {Kemmer}, {Kral}, {Lagrange}, {Launhardt}, {Matter}, {M{\'e}nard}, {Milli}, {Molli{\`e}re}, {Olofsson}, {P{\'e}rez}, {Pinilla}, {Pinte}, {Quanz}, {Schmidt}, {Udry}, {Wahhaj}, {Williams}, {Buenzli}, {Cudel}, {Dominik}, {Galicher}, {Kasper}, {Lannier}, {Mesa}, {Mouillet}, {Peretti}, {Perrot}, {Salter}, {Sissa}, {Wildi}, {Abe}, {Antichi}, {Augereau}, {Baruffolo}, {Baudoz}, {Bazzon}, {Beuzit}, {Blanchard}, {Brems}, {Buey}, {De Caprio}, {Carbillet}, {Carle}, {Cascone}, {Cheetham}, {Claudi}, {Costille}, {Delboulb{\'e}}, {Dohlen}, {Fantinel}, {Feautrier}, {Fusco}, {Giro}, {Gluck},
  {Gry}, {Hubin}, {Hugot}, {Jaquet}, {Le Mignant}, {Llored}, {Madec}, {Magnard}, {Martinez}, {Maurel}, {Meyer}, {M{\"o}ller-Nilsson}, {Moulin}, {Mugnier}, {Orign{\'e}}, {Pavlov}, {Perret}, {Petit}, {Pragt}, {Puget}, {Rabou}, {Ramos}, {Rigal}, {Rochat}, {Roelfsema}, {Rousset}, {Roux}, {Salasnich}, {Sauvage}, {Sevin}, {Soenke}, {Stadler}, {Suarez}, {Turatto}, \& {Weber}}]{Keppler_etal_2018A&A...617A..44K}
{Keppler}, M., {Benisty}, M., {M{\"u}ller}, A., {et~al.} 2018, \aap, 617, A44

\bibitem[{{Klahr} \& {Kley}(2006)}]{Klahr_Kley_2006A&A...445..747K}
{Klahr}, H. \& {Kley}, W. 2006, \aap, 445, 747

\bibitem[{{Kley}(1989)}]{Kley_1989A&A...208...98K}
{Kley}, W. 1989, \aap, 208, 98

\bibitem[{{Kolb} {et~al.}(2013){Kolb}, {Stute}, {Kley}, \& {Mignone}}]{Kolb_etal_2013A&A...559A..80K}
{Kolb}, S.~M., {Stute}, M., {Kley}, W., \& {Mignone}, A. 2013, \aap, 559, A80

\bibitem[{{Krapp} {et~al.}(2024){Krapp}, {Kratter}, {Youdin}, {Ben{\'\i}tez-Llambay}, {Masset}, \& {Armitage}}]{Krapp_etal_2024ApJ...973..153K}
{Krapp}, L., {Kratter}, K.~M., {Youdin}, A.~N., {et~al.} 2024, \apj, 973, 153

\bibitem[{{Lau} {et~al.}(2022){Lau}, {Dr{\k{a}}{\.z}kowska}, {Stammler}, {Birnstiel}, \& {Dullemond}}]{Lau_etal_2022A&A...668A.170L}
{Lau}, T. C.~H., {Dr{\k{a}}{\.z}kowska}, J., {Stammler}, S.~M., {Birnstiel}, T., \& {Dullemond}, C.~P. 2022, \aap, 668, A170

\bibitem[{{Law} {et~al.}(2022){Law}, {Crystian}, {Teague}, {{\"O}berg}, {Rich}, {Andrews}, {Bae}, {Flaherty}, {Guzm{\'a}n}, {Huang}, {Ilee}, {Kastner}, {Loomis}, {Long}, {P{\'e}rez}, {P{\'e}rez}, {Qi}, {Rosotti}, {Ru{\'\i}z-Rodr{\'\i}guez}, {Tsukagoshi}, \& {Wilner}}]{Law_etal_2022ApJ...932..114L}
{Law}, C.~J., {Crystian}, S., {Teague}, R., {et~al.} 2022, \apj, 932, 114

\bibitem[{{Levermore} \& {Pomraning}(1981)}]{Levermoe_Pomraning_1981ApJ...248..321L}
{Levermore}, C.~D. \& {Pomraning}, G.~C. 1981, \apj, 248, 321

\bibitem[{{Lodato} {et~al.}(2019){Lodato}, {Dipierro}, {Ragusa}, {Long}, {Herczeg}, {Pascucci}, {Pinilla}, {Manara}, {Tazzari}, {Liu}, {Mulders}, {Harsono}, {Boehler}, {M{\'e}nard}, {Johnstone}, {Salyk}, {van der Plas}, {Cabrit}, {Edwards}, {Fischer}, {Hendler}, {Nisini}, {Rigliaco}, {Avenhaus}, {Banzatti}, \& {Gully-Santiago}}]{Lodato_etal_2019MNRAS.486..453L}
{Lodato}, G., {Dipierro}, G., {Ragusa}, E., {et~al.} 2019, \mnras, 486, 453

\bibitem[{{Melon Fuksman} \& {Klahr}(2022)}]{Melon_Fuksman_2022ApJ...936...16M}
{Melon Fuksman}, J.~D. \& {Klahr}, H. 2022, \apj, 936, 16

\bibitem[{{Mihalas} \& {Weibel Mihalas}(1984)}]{Mihalas_WeibelMihalas_1984frh..book.....M}
{Mihalas}, D. \& {Weibel Mihalas}, B. 1984, {Foundations of radiation hydrodynamics} (Oxford University Press, New York)

\bibitem[{{Montesinos} {et~al.}(2015){Montesinos}, {Cuadra}, {Perez}, {Baruteau}, \& {Casassus}}]{Montesinos_etal_2015ApJ...806..253M}
{Montesinos}, M., {Cuadra}, J., {Perez}, S., {Baruteau}, C., \& {Casassus}, S. 2015, \apj, 806, 253

\bibitem[{{Morfill} \& {Voelk}(1984)}]{Morfill_Voelk_1984ApJ...287..371M}
{Morfill}, G.~E. \& {Voelk}, H.~J. 1984, \apj, 287, 371

\bibitem[{{Muley} {et~al.}(2024){Muley}, {Melon Fuksman}, \& {Klahr}}]{Muley_etal_2024A&A...687A.213M}
{Muley}, D., {Melon Fuksman}, J.~D., \& {Klahr}, H. 2024, \aap, 687, A213

\bibitem[{{Nakagawa} {et~al.}(1986){Nakagawa}, {Sekiya}, \& {Hayashi}}]{Nakagawa_etal_1986Icar...67..375N}
{Nakagawa}, Y., {Sekiya}, M., \& {Hayashi}, C. 1986, \icarus, 67, 375

\bibitem[{{{\"O}berg} {et~al.}(2011){{\"O}berg}, {Murray-Clay}, \& {Bergin}}]{Oberg_etal_2011ApJ...743L..16O}
{{\"O}berg}, K.~I., {Murray-Clay}, R., \& {Bergin}, E.~A. 2011, \apjl, 743, L16

\bibitem[{{Paardekooper} \& {Mellema}(2006)}]{Paardekooper_Mellema_2006A&A...459L..17P}
{Paardekooper}, S.-J. \& {Mellema}, G. 2006, \aap, 459, L17

\bibitem[{{P{\'e}rez} {et~al.}(2018){P{\'e}rez}, {Benisty}, {Andrews}, {Isella}, {Dullemond}, {Huang}, {Kurtovic}, {Guzm{\'a}n}, {Zhu}, \& {Birnstiel}}]{Perez_etal_2018ApJ...869L..50P}
{P{\'e}rez}, L.~M., {Benisty}, M., {Andrews}, S.~M., {et~al.} 2018, \apj, 869, L50

\bibitem[{{P{\'e}rez} {et~al.}(2020){P{\'e}rez}, {Casassus}, {Hales}, {Marino}, {Cheetham}, {Zurlo}, {Cieza}, {Dong}, {Alarc{\'o}n}, {Ben{\'\i}tez-Llambay}, {Fomalont}, \& {Avenhaus}}]{Perez_etal_2020ApJ...889L..24P}
{P{\'e}rez}, S., {Casassus}, S., {Hales}, A., {et~al.} 2020, \apjl, 889, L24

\bibitem[{{Perez} {et~al.}(2015){Perez}, {Dunhill}, {Casassus}, {Roman}, {Szul{\'a}gyi}, {Flores}, {Marino}, \& {Montesinos}}]{Perez_etal_2015ApJ...811L...5P}
{Perez}, S., {Dunhill}, A., {Casassus}, S., {et~al.} 2015, \apjl, 811, L5

\bibitem[{{Petrovic} {et~al.}(2024){Petrovic}, {Booth}, \& {Clarke}}]{Petrovic_etal_2024MNRAS.534.2412P}
{Petrovic}, H.~J., {Booth}, R.~A., \& {Clarke}, C.~J. 2024, \mnras, 534, 2412

\bibitem[{{Pinilla} {et~al.}(2012){Pinilla}, {Birnstiel}, {Ricci}, {Dullemond}, {Uribe}, {Testi}, \& {Natta}}]{Pinilla_etal_2012A&A...538A.114P}
{Pinilla}, P., {Birnstiel}, T., {Ricci}, L., {et~al.} 2012, \aap, 538, A114

\bibitem[{{Pinilla} \& {Youdin}(2017)}]{Pinilla_Youdin_2017ASSL..445...91P}
{Pinilla}, P. \& {Youdin}, A. 2017, in Astrophysics and Space Science Library, Vol. 445, Formation, Evolution, and Dynamics of Young Solar Systems, ed. M.~{Pessah} \& O.~{Gressel}, 91

\bibitem[{{Pinte} {et~al.}(2023){Pinte}, {Hammond}, {Price}, {Christiaens}, {Andrews}, {Chauvin}, {P{\'e}rez}, {Jorquera}, {Garg}, {Norfolk}, {Calcino}, \& {Bonnefoy}}]{Pinte_etal_2023MNRAS.526L..41P}
{Pinte}, C., {Hammond}, I., {Price}, D.~J., {et~al.} 2023, \mnras, 526, L41

\bibitem[{{Pinte} {et~al.}(2018{\natexlab{a}}){Pinte}, {M{\'e}nard}, {Duch{\^e}ne}, {Hill}, {Dent}, {Woitke}, {Maret}, {van der Plas}, {Hales}, {Kamp}, {Thi}, {de Gregorio-Monsalvo}, {Rab}, {Quanz}, {Avenhaus}, {Carmona}, \& {Casassus}}]{Pinte_etal_2018A&A...609A..47P}
{Pinte}, C., {M{\'e}nard}, F., {Duch{\^e}ne}, G., {et~al.} 2018{\natexlab{a}}, \aap, 609, A47

\bibitem[{{Pinte} {et~al.}(2018{\natexlab{b}}){Pinte}, {Price}, {M{\'e}nard}, {Duch{\^e}ne}, {Dent}, {Hill}, {de Gregorio-Monsalvo}, {Hales}, \& {Mentiplay}}]{Pinte_etal_2018ApJ...860L..13P}
{Pinte}, C., {Price}, D.~J., {M{\'e}nard}, F., {et~al.} 2018{\natexlab{b}}, \apjl, 860, L13

\bibitem[{{Pinte} {et~al.}(2019){Pinte}, {van der Plas}, {M{\'e}nard}, {Price}, {Christiaens}, {Hill}, {Mentiplay}, {Ginski}, {Choquet}, {Boehler}, {Duch{\^e}ne}, {Perez}, \& {Casassus}}]{Pinte_etal_2019NatAs...3.1109P}
{Pinte}, C., {van der Plas}, G., {M{\'e}nard}, F., {et~al.} 2019, Nature Astronomy, 3, 1109

\bibitem[{{Qi} {et~al.}(2013){Qi}, {{\"O}berg}, {Wilner}, {D'Alessio}, {Bergin}, {Andrews}, {Blake}, {Hogerheijde}, \& {van Dishoeck}}]{Qi_etal_2013Sci...341..630Q}
{Qi}, C., {{\"O}berg}, K.~I., {Wilner}, D.~J., {et~al.} 2013, Science, 341, 630

\bibitem[{{Qi} \& {Wilner}(2024)}]{Qi_Wilner_2024ApJ...977...60Q}
{Qi}, C. \& {Wilner}, D.~J. 2024, \apj, 977, 60

\bibitem[{{Remijan} {et~al.}(2019){Remijan}, {Biggs}, {Cortes}, {Dent}, {Di Franceso}, {Fomalont}, {Hales}, {Kameno}, {Mason}, {Philips}, {Saini}, {Vila Vilaro}, \& {Villard}}]{2019athb.rept.....R}
{Remijan}, A., {Biggs}, A., {Cortes}, P.~A., {et~al.} 2019, {ALMA Technical Handbook,ALMA Doc. 7.3, ver. 1.1}, 2019, ALMA Technical Handbook,ALMA Doc. 7.3, ver. 1.1ISBN 978-3-923524-66-2

\bibitem[{{Sagynbayeva} {et~al.}(2024){Sagynbayeva}, {Li}, {Kuznetsova}, {Zhu}, {Jiang}, \& {Armitage}}]{Sagynbaeva_etal_2024arXiv241014896S}
{Sagynbayeva}, S., {Li}, R., {Kuznetsova}, A., {et~al.} 2024, arXiv e-prints, arXiv:2410.14896

\bibitem[{{Sch{\"o}ier} {et~al.}(2005){Sch{\"o}ier}, {van der Tak}, {van Dishoeck}, \& {Black}}]{Schoier_etal_2005A&A...432..369S}
{Sch{\"o}ier}, F.~L., {van der Tak}, F.~F.~S., {van Dishoeck}, E.~F., \& {Black}, J.~H. 2005, \aap, 432, 369

\bibitem[{{Schwarz} {et~al.}(2016){Schwarz}, {Bergin}, {Cleeves}, {Blake}, {Zhang}, {{\"O}berg}, {van Dishoeck}, \& {Qi}}]{Schwarz_etal_2016ApJ...823...91S}
{Schwarz}, K.~R., {Bergin}, E.~A., {Cleeves}, L.~I., {et~al.} 2016, \apj, 823, 91

\bibitem[{{Shakura} \& {Sunyaev}(1973)}]{Shakura_Sunyaev_1973A&A....24..337S}
{Shakura}, N.~I. \& {Sunyaev}, R.~A. 1973, \aap, 500, 33

\bibitem[{{Stone} \& {Norman}(1992)}]{Stone_Norman_1992ApJS...80..753S}
{Stone}, J.~M. \& {Norman}, M.~L. 1992, \apjs, 80, 753

\bibitem[{{Szul{\'a}gyi}(2017)}]{Szulagyi_2017ApJ...842..103S}
{Szul{\'a}gyi}, J. 2017, \apj, 842, 103

\bibitem[{{Szul{\'a}gyi} {et~al.}(2016){Szul{\'a}gyi}, {Masset}, {Lega}, {Crida}, {Morbidelli}, \& {Guillot}}]{Szulagyi_etal_2016MNRAS.460.2853S}
{Szul{\'a}gyi}, J., {Masset}, F., {Lega}, E., {et~al.} 2016, \mnras, 460, 2853

\bibitem[{{Szul{\'a}gyi} {et~al.}(2018){Szul{\'a}gyi}, {Plas}, {Meyer}, {Pohl}, {Quanz}, {Mayer}, {Daemgen}, \& {Tamburello}}]{Szulagyi_etal_2018MNRAS.473.3573S}
{Szul{\'a}gyi}, J., {Plas}, G. v.~d., {Meyer}, M.~R., {et~al.} 2018, \mnras, 473, 3573

\bibitem[{{Teague} {et~al.}(2018){Teague}, {Bae}, {Bergin}, {Birnstiel}, \& {Foreman-Mackey}}]{Teague2018ApJ...860L..12T}
{Teague}, R., {Bae}, J., {Bergin}, E.~A., {Birnstiel}, T., \& {Foreman-Mackey}, D. 2018, \apjl, 860, L12

\bibitem[{{Ueda} {et~al.}(2024){Ueda}, {Tazaki}, {Okuzumi}, {Flock}, \& {Sudarshan}}]{Ueda_etal_2024NatAs...8.1148U}
{Ueda}, T., {Tazaki}, R., {Okuzumi}, S., {Flock}, M., \& {Sudarshan}, P. 2024, Nature Astronomy, 8, 1148

\bibitem[{{Wagner} {et~al.}(2018){Wagner}, {Follete}, {Close}, {Apai}, {Gibbs}, {Keppler}, {M{\"u}ller}, {Henning}, {Kasper}, {Wu}, {Long}, {Males}, {Morzinski}, \& {McClure}}]{Wagner_etal_2018ApJ...863L...8W}
{Wagner}, K., {Follete}, K.~B., {Close}, L.~M., {et~al.} 2018, \apjl, 863, L8

\bibitem[{{Wagner} {et~al.}(2023){Wagner}, {Stone}, {Skemer}, {Ertel}, {Dong}, {Apai}, {Spalding}, {Leisenring}, {Sitko}, {Kratter}, {Barman}, {Marley}, {Miles}, {Boccaletti}, {Assani}, {Bayyari}, {Uyama}, {Woodward}, {Hinz}, {Briesemeister}, {Lawson}, {M{\'e}nard}, {Pantin}, {Russell}, {Skrutskie}, \& {Wisniewski}}]{Wagner_etal_2023NatAs...7.1208W}
{Wagner}, K., {Stone}, J., {Skemer}, A., {et~al.} 2023, Nature Astronomy, 7, 1208

\bibitem[{{Weber} {et~al.}(2019){Weber}, {P{\'e}rez}, {Ben{\'\i}tez-Llambay}, {Gressel}, {Casassus}, \& {Krapp}}]{Weber_etal_2019ApJ...884..178W}
{Weber}, P., {P{\'e}rez}, S., {Ben{\'\i}tez-Llambay}, P., {et~al.} 2019, \apj, 884, 178

\bibitem[{{Wilson} \& {Rood}(1994)}]{Wilson_Rood_1994ARA&A..32..191W}
{Wilson}, T.~L. \& {Rood}, R. 1994, \araa, 32, 191

\bibitem[{{Wolf} \& {D'Angelo}(2005)}]{Wolf_DAngelo_2005ApJ...619.1114W}
{Wolf}, S. \& {D'Angelo}, G. 2005, \apj, 619, 1114

\bibitem[{{Zhang} {et~al.}(2018){Zhang}, {Zhu}, {Huang}, {Guzm{\'a}n}, {Andrews}, {Birnstiel}, {Dullemond}, {Carpenter}, {Isella}, \& {P{\'e}rez}}]{Zhang_etal_2018ApJ...869L..47Z}
{Zhang}, S., {Zhu}, Z., {Huang}, J., {et~al.} 2018, \apj, 869, L47

\bibitem[{Zhang {et~al.}(2023)Zhang, Zhu, Ueda, Kataoka, Sierra, Carrasco-González, \& Macías}]{Zhang_2023}
Zhang, S., Zhu, Z., Ueda, T., {et~al.} 2023, The Astrophysical Journal, 953, 96

\bibitem[{{Zhou} {et~al.}(2025){Zhou}, {Bowler}, {Sanghi}, {Marleau}, {Takasao}, {Aoyama}, {Hasegawa}, {Thanathibodee}, {Uyama}, {Hashimoto}, {Wagner}, {Calvet}, {Demars}, {Wu}, {Biddle}, {Haffert}, \& {Bryan}}]{Zhou_etal_2025ApJ...980L..39Z}
{Zhou}, Y., {Bowler}, B.~P., {Sanghi}, A., {et~al.} 2025, \apjl, 980, L39

\bibitem[{{Zhou} {et~al.}(2021){Zhou}, {Bowler}, {Wagner}, {Schneider}, {Apai}, {Kraus}, {Close}, {Herczeg}, \& {Fang}}]{Zhou_etal_2021AJ....161..244Z}
{Zhou}, Y., {Bowler}, B.~P., {Wagner}, K.~R., {et~al.} 2021, \aj, 161, 244

\bibitem[{{Zhu} {et~al.}(2018){Zhu}, {Andrews}, \& {Isella}}]{zhu2018MNRAS.479.1850Z}
{Zhu}, Z., {Andrews}, S.~M., \& {Isella}, A. 2018, \mnras, 479, 1850

\bibitem[{{Zhu} {et~al.}(2012){Zhu}, {Nelson}, {Dong}, {Espaillat}, \& {Hartmann}}]{Zhu_etal_2012ApJ...755....6Z}
{Zhu}, Z., {Nelson}, R.~P., {Dong}, R., {Espaillat}, C., \& {Hartmann}, L. 2012, \apj, 755, 6

\bibitem[{{Ziampras} {et~al.}(2025){Ziampras}, {Sudarshan}, {Dullemond}, {Flock}, {Berta}, {Nelson}, \& {Mignone}}]{Ziampras_etal_2025MNRAS.536.3322Z}
{Ziampras}, A., {Sudarshan}, P., {Dullemond}, C.~P., {et~al.} 2025, \mnras, 536, 3322

\end{thebibliography}

\begin{appendix}

\section{Dust depletion analysis}
\label{sec:dust_depletion}

\begin{figure*}[!b]
\centering
\includegraphics[width=0.9\linewidth]{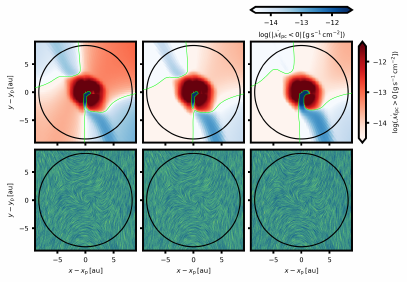}
\caption{\emph{Top:} Midplane radial mass flux of large dust grains 
in a planetocentric coordinate frame. The composite color gradient
shows $\log{\dot{\mathcal{M}}_{\mathrm{pc}}}$ wherever $\dot{\mathcal{M}}_{\mathrm{pc}}>0$ (red; flux towards the planet) and $\log{|\dot{\mathcal{M}}_{\mathrm{pc}}|}$ wherever $\dot{\mathcal{M}}_{\mathrm{pc}}<0$ (blue; flux away from the planet). 
The underlying planetocentric mesh was only used for the purpose of calculating $\dot{\mathcal{M}}_{\mathrm{pc}}$; it does not correspond to the mesh used in our hydrodynamic simulations. The black circle marks the Hill radius. The green isocontour is where $\dot{\mathcal{M}}_{\mathrm{pc}}=0$.
\emph{Bottom:} Flow topology of large dust grains obtained with the 
line integral convolution (LIC) method. Individual \emph{columns} (\emph{left} to \emph{right}, respectively)
correspond to simulation times $t=375$, $400$, and $425$ orbits.}
\label{fig:massflux}
\end{figure*}

\FloatBarrier

In Sect.~\ref{sec:outflows}, we suggested that the main pathway
through which large dust grains leave the circumplanetary environment
is via the horseshoe flow. 
As an additional confirmation, here we analyze the mass flux of large dust grains at
various simulation times.
To this end, we first constructed a planetocentric spherical grid and mapped $(\rho\gas,\vec{v})$ onto it using trilinear interpolation. Second, we calculated the radial mass flux of large dust defined in the new planetocentric coordinates as
\begin{equation}
\dot{\mathcal{M}}_{\mathrm{pc}} = - \rho\bg u_{\mathrm{pc},r} \, ,
\label{eq:mass_flux}
\end{equation}
where $u_{\mathrm{pc},r}$ is the radial velocity of large dust grains with respect to the planet. The negative sign in Eq.~\ref{eq:mass_flux} implies that $\dot{\mathcal{M}}_{\mathrm{pc}}>0$ when large dust is moving towards the planet ($u_{\mathrm{pc},r}<0$) and $\dot{\mathcal{M}}_{\mathrm{pc}}<0$ when dust is moving away from the planet ($u_{\mathrm{pc},r}>0$).

Fig.~\ref{fig:massflux} (top row) shows the time evolution of $\dot{\mathcal{M}}_{\mathrm{pc}}$. Despite the fact that the radial mass flux of dust is headed towards the planet in the majority of the inner Hill sphere, there is a persistent outflow channel trailing the planet location (near $x=x_{\mathrm{p}}$ and $y<y_{\mathrm{p}}$) and connecting to the rear horseshoe flow (compare with the flow topology shown in the bottom row of Fig.~\ref{fig:massflux}). The magnitude of this outflow does not seem to diminish over the displayed interval $t=375$--$425\,\mathrm{orbits}$, contrary to the rest of the flow across the Hill radius which decreases over time
as a natural consequence of pressure bump filtering. This confirms the dominant role of the rear horseshoe flow in depleting the dust accumulation surrounding the planet. Turbulent dust diffusion, which is included in our numerical model, likely aids the process because it keeps the central dust accumulation spread-out rather than sharply centrally peaked.

Finally, we point out that we also analyzed $\dot{\mathcal{M}}_{\mathrm{pc}}$ in the meridional plane, motivated by the presence of polar outflows in Fig.~\ref{fig:dustplume_vert}, but did not find an out-flowing mass flux of a magnitude comparable to that shown in Fig.~\ref{fig:massflux}.

\section{Additional simulations}
\label{sec:additional}

\begin{figure*}
    \centering
    \includegraphics[width=0.4\linewidth]{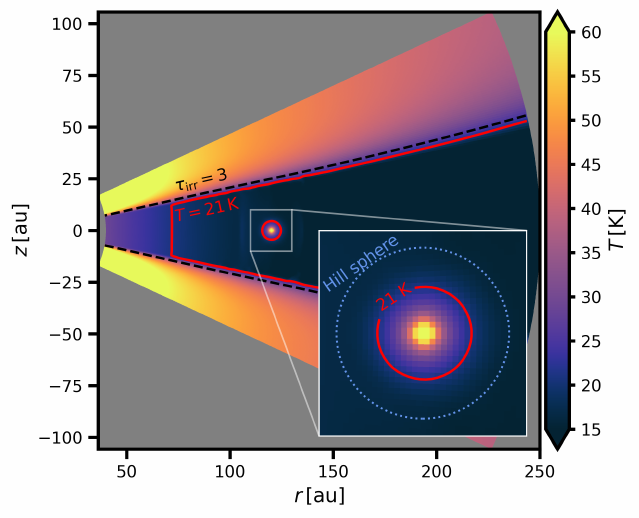}
    \includegraphics[width=0.4\linewidth]{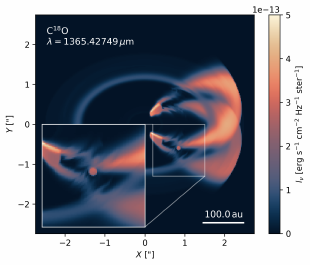}
    \includegraphics[width=0.4\linewidth]{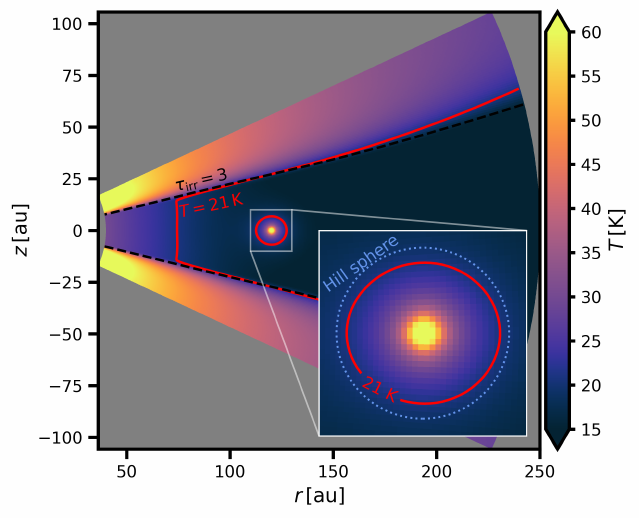}
    \includegraphics[width=0.4\linewidth]{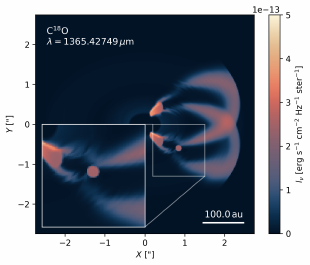}
    \includegraphics[width=0.4\linewidth]{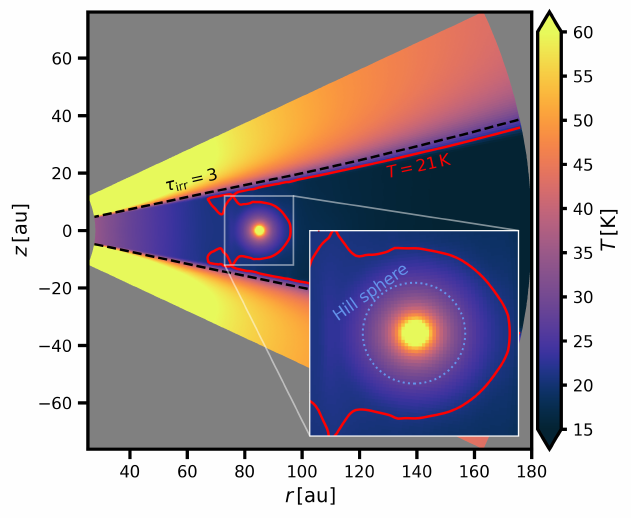}
    \includegraphics[width=0.4\linewidth]{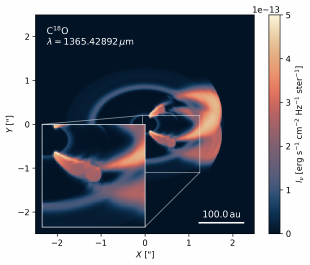}
    \caption{Meridional profiles of the disk temperature
    (\emph{left}) and synthetic channel maps of C$^{18}$O emission
    (\emph{right}) for simulations S1 (\emph{top}), S2 (\emph{middle}), and S3 (\emph{bottom}). 
    Please note that the planet is shifted to $85\,\mathrm{au}$ in the \emph{bottom row}.
    Therefore, the radial extent of the simulated disk,
    the wavelength of the synthetic image, and the centering of the inset
    differ from simulations S1 and S2.}
    \label{fig:s1s2s3}
\end{figure*}

\begin{table*}[]
    \caption{Summary of differences between additional simulations
    discussed in Appendix~\ref{sec:additional} and the nominal simulation.}
    \centering
    \begin{tabular}{lccccc}
    \hline\hline
    Simulation & Scold & S1 & S2 & S3 & S4 \\
    \hline
    Mass doubling time $t_{\mathrm{acc}}$\tablefootmark{(a)} & N/A & $0.5\,\mathrm{Myr}$ & $0.5\,\mathrm{Myr}$ & -- & -- \\
    Opacity law & -- & -- & uniform\tablefootmark{(b)} & -- & -- \\
    Planet's orbital distance $r_{\mathrm{p}}$\tablefootmark{(a)} & -- & -- & -- & $85\,\mathrm{au}$ & -- \\
    Inner radial boundary\tablefootmark{(a)} & -- & -- & -- & $28\,\mathrm{au}$ & -- \\
    Outer radial boundary\tablefootmark{(a)} & -- & -- & -- & $180\,\mathrm{au}$ & -- \\
    Planet mass $M_{\mathrm{p}}$\tablefootmark{(a)} & -- & -- & -- & -- & $20\,M_{\oplus}$ \\
    Smoothing length $r_{\mathrm{sm}}$\tablefootmark{(a)} & -- & -- & -- & -- & $0.5\,R_{\mathrm{H}}$ \\
    No. of orbits to introduce $M_{\mathrm{p}}$\tablefootmark{(a)} & -- & -- & -- & -- & 5 \\
    No. of simulated orbits\tablefootmark{(a)} & -- & -- & -- & -- & 25 \\
    Putative gap amplitude $A_{\mathrm{g}}$ & N/A & N/A & N/A & N/A & $5$ \\
    Putative gap centre $r_{\mathrm{g}}$ & N/A & N/A & N/A & N/A & $80\,\mathrm{au}$ \\
    Putative gap width $w_{\mathrm{g}}$ & N/A & N/A & N/A & N/A & $15\,\mathrm{au}$ \\    
    \hline
    \end{tabular}
    \tablefoot{Dashes imply that a parameter keeps the nominal value. 
    When a parameter is not applicable (not used in the model), we write `N/A'.
    \tablefoottext{a}{Applies to the final simulation stage with an embedded planet. The time span of the hydrodynamic relaxation stage is 800 orbits for simulation S4 and 200 orbits otherwise.}
    \tablefoottext{b}{When using the primitive uniform opacity model, we assume that
    the Planck and Rosseland opacities are identical and equal to $1\,\mathrm{cm}^{2}\,\mathrm{g}^{-1}$
    while the opacity to stellar irradiation amounts to $3\,\mathrm{cm}^{2}\,\mathrm{g}^{-1}$
    (all values being expressed per gram of gas).}
    }
    \label{tab:params_additional}
\end{table*}

The nominal simulation discussed in the main body of the paper
is based on several conceptual and parametric choices that could be perceived 
as rather extreme. These are mainly
\begin{enumerate}
    \item a relatively strong energy output of the planet ($T_{\mathrm{p,e}}\simeq4000\,\mathrm{K}$) and,
    \item the very existence of a massive Jupiter-like planet at a large orbital separation ($r_{\mathrm{p}}=120\,\mathrm{au}$).
\end{enumerate}
In this Appendix, we present four additional simulations that 
we selected to demonstrate
how the CO bubble changes when points (1.) and (2.) are modified.
Additionally, we compare the nominal simulation to a 
case of a nonaccreting planet ($L_{\mathrm{p}}=0$) in order
to discuss its ability to form a CPD.
A brief overview of the additional simulations is provided in Table~\ref{tab:params_additional}
while a detailed explanation is given in the following.

\subsection{Simulations S1 and S2}
\label{sec:s1s2}

Regarding point (1.), some level of degeneracy is expected between the luminosity
of the planet and the opacity law, with the latter controlling the cooling rate of the planet's vicinity.
On one hand, when decreasing $L_{\mathrm{p}}$, the size of the bubble should shrink
as the local heating source for the gas weakens.
On the other hand, assuming a different opacity law
might help spreading the hot spot over a larger distance via radiative diffusion
and thus compensate for the decrease in $L_{\mathrm{p}}$.
With this in mind, we conducted two additional simulations.
Simulation labeled S1 differs from the nominal simulation in the mass doubling
time of the planet, which is increased to $t_{\mathrm{acc}}=0.5\,\mathrm{Myr}$.
The accretion luminosity and effective temperature decrease to $L_{\mathrm{p}}\simeq5\times10^{-4}\,L_{\odot}$
and $T_{\mathrm{p,e}}\simeq2650\,\mathrm{K}$, respectively.
Simulation labeled S2 has the same less efficient accretion as S1
but the opacity is set to uniform values
$\kappa_{\mathrm{P}}=\kappa_{\mathrm{R}}=1\,\mathrm{cm}^{2}\,\mathrm{g}^{-1}$
and $\kappa_{\star}=3\,\mathrm{cm}^{2}\,\mathrm{g}^{-1}$ per gram of gas.
We evolve only gas in simulation S2 because the influence
of large grains on the opacity is not considered.
Additionally, it was necessary to perform the hydrostatic and hydrodynamic relaxations for simulation S2 again because
the change in the opacity leads to a slightly
different equilibrium disk profile.

The resulting temperature profiles and C$^{18}$O channel
maps for simulations S1 and S2 are shown in Fig.~\ref{fig:s1s2s3} (top and middle rows, respectively). In simulation S1, the diameter of the bubble 
shrinks to $45\%$ compared to our nominal simulation.
However, the shrinkage can indeed 
be thwarted by a change in the opacity law, as resulting
from simulation S2 in which the bubble diameter retains
$73\%$ of the nominal extent.

The behavior demonstrated by simulations S1 and S2 
implies that the extent of the bubble critically depends 
not only on the accretion luminosity, but also 
on the optical thickness of the circumplanetary region
to escaping thermal radiation. The optical thickness can 
be changed not only due to the opacity but also due to 
a change in the dust content (note that the opacity in simulation
S1 is tied to the abundance of large dust grains, which
becomes low close to the planet). Therefore,
future models should strive for a detailed understanding
of dust accumulation close to the planet across a wide range
of grain sizes.

\begin{figure}
    \centering
    \includegraphics[width=0.9\columnwidth]{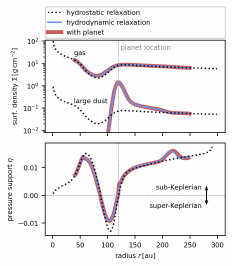}
    \caption{\emph{Top:} As in the upper panel of Fig.~\ref{fig:M1_profiles}, but for simulation S4 in which
    we introduce a putative gap in the gas disk before inserting a
    subthermal-mass planet.
    The planet is introduced and evolved inside the 
    dusty ring occurring at the outer edge of the putative gap.
    \emph{Bottom:} Radial profile of the pressure support parameter $\eta=-1/2(H/r)^{2}\partial\log{P}/\partial\log{r}$ \citep[e.g.][]{Nakagawa_etal_1986Icar...67..375N,Bitsch_etal_2018A&A...612A..30B}
    which delimits regions of sub- and super-Keplerian rotation.}
    \label{fig:s4_profiles}
\end{figure}

\subsection{Simulation S3}
\label{sec:s3}

Returning to point (2.), there are essentially two possibilities
to modify our nominal assumption.
First, perhaps the likelihood of forming a Jupiter-mass planet
is larger closer to the star. Our nominal simulation does allow for some wiggle room as the planet can be shifted inwards
and still end up outside the CO snowline.
Therefore, we performed simulation S3 in which the Jupiter-mass planet
was placed at $r_{\mathrm{pl}}=85\,\mathrm{au}$ and 
the radial span of the domain was rescaled accordingly,
ranging from $28$ to $180\,\mathrm{au}$.
Second, it might be worth exploring the accretion feedback
from subthermal-mass planets, which is done in Sect.~\ref{sec:s4}.

The results of simulation S3 are displayed in Fig.~\ref{fig:s1s2s3} (bottom row).
The size of the bubble is similar to the nominal simulation, which is not surprising
since we only changed the planet's orbital radius. However, due to its proximity
to the CO snowline, the bubble partially merges with it. Subsequently, when looking 
at the synthetic image, the bubble appears to be attached to one of the wedges
which fill out the space between the dragonfly wings.

\subsection{Simulation S4}
\label{sec:s4}

Recent theoretical works suggest that pressure bumps with dusty rings can promote
fast growth of low-mass embryos until they reach masses
of giant-planet cores. For instance, 
considering pressure bumps at $\sim$$70\,\mathrm{au}$
distances, 
\cite{Lau_etal_2022A&A...668A.170L} and \cite{Jiang_Ormel_2023MNRAS.518.3877J} showed
that planets growing by pebble and planetesimal accretion reach $M_{\mathrm{p}}\sim20\,M_{\oplus}$
on a timescale $\sim$$0.1\,\mathrm{Myr}$.
Therefore, instead of a Jupiter-mass planet, one could 
argue for subthermal-mass planets to be more easier to assemble
in the outermost disk regions, while also exhibiting nonzero
accretion luminosities.

Motivated by these findings, we designed simulation labeled S4
to study the CO bubble around a $M_{\mathrm{p}}=20\,M_{\oplus}$
planet, having the mass doubling time $t_{\mathrm{acc}}=0.1\,\mathrm{Myr}$ and being positioned within a dust-loaded
pressure bump. To place a putative bump in the disk
before the planet is introduced, we modify Eq.~(\ref{eq:surfdens})
by carving a Gaussian gap
in the surface density profile\citep{Pinilla_etal_2012A&A...538A.114P,Dullemond_etal_2018ApJ...869L..46D,Chrenko_Chametla_2023MNRAS.524.2705C}:
\begin{equation}
    \Sigma'\gas = \Sigma\gas \left[1+(A_{\mathrm{g}}-1)\exp{\left(-\frac{(r-r_{\mathrm{g}})^{2}}{2w_{\mathrm{g}}^{2}}\right)}\right]^{-1} \, ,
    \label{eq:bump}
\end{equation}
where $A_{\mathrm{g}}$ is the amplitude of the gap, $r_{\mathrm{g}}$ is the radial centre
of the perturbation, and $w_{\mathrm{g}}$ sets the width of the Gaussian. We note
that an opposite perturbation (i.e. a peak) in the disk viscosity
\begin{equation}
    \nu' = \nu \left[1+(A_{\mathrm{g}}-1)\exp{\left(-\frac{(r-r_{\mathrm{g}})^{2}}{2w_{\mathrm{g}}^{2}}\right)}\right] \, ,
    \label{eq:viscobump}
\end{equation}
has to be considered to ensure stability of the gap profile 
on a viscous time scale. At the outer edge of the gap,
a pressure bump forms (similarly to a planet-induced gap)
and becomes loaded by large dust grains (Fig.~\ref{fig:s4_profiles}).

\begin{figure*}[]
    \centering
    \includegraphics[width=0.4\linewidth]{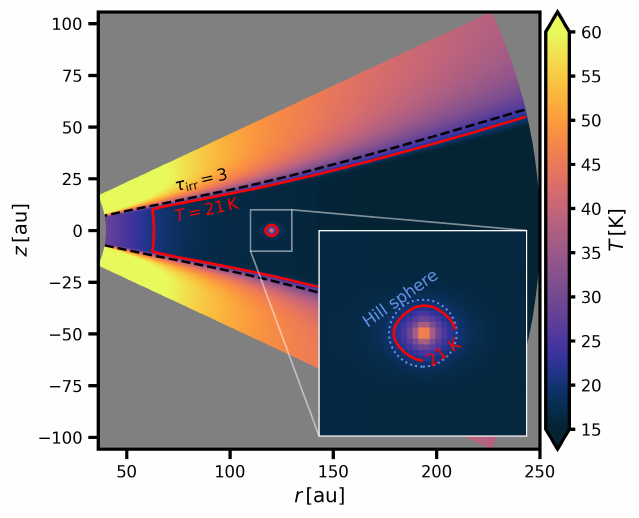}
    \includegraphics[width=0.4\linewidth]{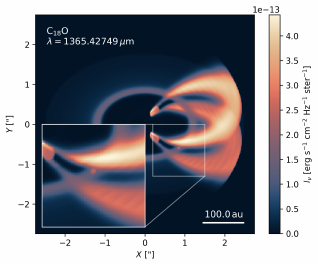}
    \caption{Meridional profiles of the disk temperature
    (\emph{left}) and synthetic channel maps of C$^{18}$O emission
    (\emph{right}) for simulation S4 in which a subthermal-mass
    planet evolves in a dusty ring.}
    \label{fig:s4}
\end{figure*}

\begin{figure}[!h]
    \centering
    \includegraphics[width=0.95\linewidth]{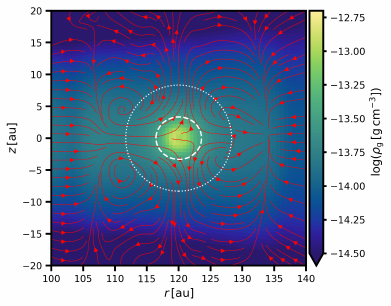}
    \includegraphics[width=0.95\linewidth]{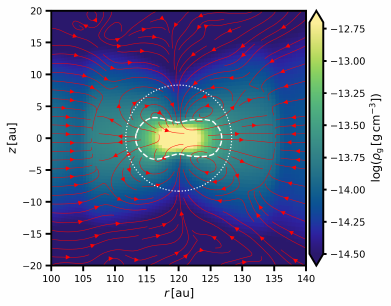}
    \caption{Logarithm of the vertical gas density distribution in the vicinity of the planet 
    at $t=375$ orbits (the same orbital time as in Fig.~\ref{fig:dustplume}). Overlaid
    are the streamlines of the gaseous meridional flow (red arrows), the Hill sphere (white dotted
    circle), and an arbitrarily chosen isosurface of $\log{(\rho_{\mathrm{g}}/\mathrm{g}\,\mathrm{cm}^{-3})}=-13.5$
    (white dashed curve). \emph{Top:} nominal simulation. \emph{Bottom:} simulation Scold
    with a zero-luminosity planet.}
    \label{fig:hot_vs_cold}
\end{figure}

Compared to the nominal simulation, we ran the
hydrodynamic relaxation stage for 800 orbits.
At the end of the relaxation, the peak increase
of $\Sigma\bg$ in the pressure bump was by a factor of $20$
with respect to initial conditions (see Fig.~\ref{fig:s4_profiles}).
The main stage with the planet was shortened to 25 orbits
because the given planet mass perturbs the disk only weakly.
During the simulation, we removed some large dust from the immediate
vicinity of the planet in order to keep the dust-to-gas ratio below $90\%$. This removed dust was not added to the mass of the planet,
nor was it linked to the accretion luminosity (we kept it parametric).

Looking at Fig.~\ref{fig:s4}, the diameter of the CO bubble is quite small,
dropping to $30\%$ compared to the nominal case.
This is because the accretion luminosity in simulation S4 is only $L_{\mathrm{p}}\simeq2\times10^{-5}\,L_{\odot}$.
Still, the diameter of the bubble is not completely negligible owing to the surrounding dense dusty ring which boosts local optical depths and
helps spreading the temperature perturbation further away from the planet.

We attempted the same kinematic analysis as in Sect.~\ref{sec:kinematics}.
The result, however, is negative because the bubble ends up at
the thermal noise level and cannot be localized as a significant
point source residual. This holds even when assuming a $\sim$$7\,\mathrm{d}$
long ALMA integration. Future work is thus needed in order to determine
the minimum accretion rate of low-mass planets necessary to produce a detectable CO bubble.

\subsection{Simulation Scold}
\label{sec:scold}

To gain more insight into the structure of the circumplanetary environment,
it is instructive to compare the nominal simulation with a case of a zero-luminosity
planet (simulation Scold). Such a comparison is given in Fig.~\ref{fig:hot_vs_cold}
in terms of the vertical gas density and flow.
When the planet is cold, the circumplanetary streamlines exhibit a polar inflow and equatorial
outflow \citep[e.g.][]{Ayliffe_Bate_2009MNRAS.397..657A,Szulagyi_etal_2016MNRAS.460.2853S}.
Nevertheless, the level of vertical gas compression and deviation from a quasi-spherical 
envelope structure is only moderate. This is related to the fact that the ratio
of the planet mass to the local thermal mass ($M_{\mathrm{th}}=h^{3}M_{\star}$)
is close to unity, $M_{\mathrm{p}}/M_{\mathrm{th}}\simeq1.4$. For planets at the verge
of the thermal mass, \cite{Sagynbaeva_etal_2024arXiv241014896S} recently showed that
the level of circumplanetary `diskiness' indeed remains only moderate
(their figure 6 for $h=0.1$ and $M_{\mathrm{p}}=1\,M_{\mathrm{Jup}}$ can serve as a useful comparison).
Their assertion that full-fledged CPDs are unlikely to form around 
Jupiter-mass protoplanets at large orbital separations
(due to the increase of the disk's aspect ratio)
matches our findings.

When the nominal accretion luminosity is considered,
the local heat release provides additional pressure support 
counteracting the vertical compression \citep[see also][]{Muley_etal_2024A&A...687A.213M}
and the planet is left with a quasi-spherical envelope of lowered gas density.
Moreover, the polar flow topology becomes reversed compared to the zero-luminosity planet,
as already mentioned in Sect.~\ref{sec:outflows}.

\end{appendix}

\end{document}